\documentclass{ws-ijmpd}
\usepackage[super,compress]{cite}
\usepackage{hyperref}
\hypersetup{colorlinks,urlcolor=black,citecolor=black,linkcolor=black,filecolor=black}
\usepackage{breakurl}
\begin{document}

\markboth{M. R. Feldman \& J. D. Anderson}
{On the possible onset of the Pioneer anomaly}

%
\catchline{}{}{}{}{}
%

\title{ON THE POSSIBLE ONSET OF THE PIONEER ANOMALY }

\author{MICHAEL R. FELDMAN}

\address{Southwest Research Institute, 6220 Culebra Road\\
San Antonio, TX 78238,
United States of America\\
mrf@m--y.us}

\author{JOHN D. ANDERSON}

\address{Southwest Research Institute, 6220 Culebra Road\\
San Antonio, TX 78238,
United States of America\\
jdandy@earthlink.net}

\maketitle

\begin{history}
\received{Day Month Year}
\revised{Day Month Year}
\end{history}

\begin{abstract}
We explore the possibility that the observed onset of the Pioneer anomaly after Saturn encounter by Pioneer 11 is not necessarily due to mismodeling of solar radiation pressure but instead reflects a physically relevant characteristic of the anomaly itself. We employ the principles of a recently proposed cosmological model termed ``the theory of inertial centers" along with an understanding of the fundamental assumptions taken by the Deep Space Network (DSN) to attempt to model this sudden onset. Due to an ambiguity that arises from the difference in the DSN definition of expected light-time with light-time according to the theory of inertial centers, we are forced to adopt a seemingly arbitrary convention to relate DSN-assumed clock-rates to physical clock-rates for this model. We offer a possible reason for adopting the convention employed in our analysis; however, we remain skeptical. Nevertheless, with this convention, one finds that this theory is able to replicate the previously reported Hubble-like behavior of the ``clock acceleration" for the Pioneer anomaly as well as the sudden onset of the anomalous acceleration after Pioneer 11 Saturn encounter. While oscillatory behavior with a yearly period is also predicted for the anomalous clock accelerations of both Pioneer 10 and Pioneer 11, the predicted amplitude is an order of magnitude too small when compared with that reported for Pioneer 10.
\end{abstract}


\ccode{PACS numbers: 95.10.Jk, 06.30.Ft, 98.80.-k}


\section{Introduction}	

First reported in Ref.~\refcite{reportPioneer} and detailed in Ref.~\refcite{Anderson}, the Pioneer anomaly refers to an unmodeled, blueshifted frequency drift with near uniform rate of change observed in the DSN residuals for both Pioneer 10 and Pioneer 11 spacecrafts with behavior in the latter portion of their cruise phases out of the Solar System resembling that of an additional near-constant acceleration in a yet to be unambiguously determined direction back toward the Solar System.\cite{TuryshevToth} While there have been many suggestions and avenues of exploration for what could possibly account for this discrepancy,\cite{Anderson, TuryshevToth} after a recent analysis of an extended set of DSN frequency data, there appears to be significant support for the origin of this anomaly as a recoil force associated with an anisotropic emission of thermal radiation off the spacecrafts due to the waste heat emitted by the radioisotope thermoelectric generators (RTGs) onboard.\cite{ThermalPioneer2012, Francisco, Modenini2014, Bertolami2012, Rievers2010, Rievers2009, Rievers_ModelThermal, Bertolami2010, Bertolami2008} Nevertheless, the reported onset of the anomaly after Saturn encounter by Pioneer 11 (see e.g. Refs.~\citen{Nieto, NietoAnderson, Iorio_JupiterSaturn} and \refcite{Iorio_orbeffects}) remains an open source of concern after the same analysis of the extended data set appeared to be consistent with the possibility of this behavior, although taken to likely be the result of mismodeling of solar radiation pressure.\cite{ThermalPioneer2011} Furthermore, other works independent of articles concerned with a thermal explanation effectively showed that the Pioneer anomaly as an extra {\it physical} acceleration directed towards the Sun acting upon all Solar System bodies would be inconsistent with the observed motion of planets, spacecrafts, and other celestial bodies.\cite{Iorio_Giudice, Fienga_gravtests, Standish_testing, Standish, Iorio_orbeffects, Iorio_lensethirringPioneer, Iorio_NeptunePioneer, Iorio_velocitydepend, Iorio_Pioneergravorigin, Iorio_JupiterSaturn, Iorio_oort, Page_HowWell, Varieschi2012, Page_MinorPlanets, Page_WeakLimit, Wallin_TestinggravOutersolar, Tangen2007} It would appear from these analyses that the Pioneer anomaly, if caused by physics not modeled by the DSN and not simply an unaccounted-for acceleration due to the intrinsic structure of the spacecrafts, would have to affect only the Doppler residuals calculated by the DSN and not the actual paths of our Solar System objects relative to one another. We will emphasize this point later in our discussion and argue that, under our proposed mechanism, objects in the Solar System will continue to orbit the Sun in the same manner as predicted by general relativity and expected by the DSN. While there have been several other documented anomalies in the Solar System that should be mentioned (reviewed in Refs.~\refcite{Anderson2010} and \refcite{Iorio2015}), such as the flyby anomaly,\cite{Gravassist} the secular increase in the astronomical unit (AU),\cite{Krasinsky2004} an anomalous secular eccentricity variation in the Moon's orbit \cite{Williams_lunarcore}, an alleged orbital anomaly reported in Saturn's perhelion which seemed to disappear upon further analysis with Cassini data,\cite{Iorio_perihelionSaturn, Acedo2014, Hees2014} and possible anomalous behavior reported in Jupiter's perhelion albeit with a low level of statistical significance,\cite{Pitjev_Pitjeva_constraintsdarkmatter, Pitjev_Pitjeva_Reldarkmatter} our concern for the present article remains solely on the Pioneer anomaly.

Our approach in this work is to explore the possibility that the reported onset is a physical characteristic of the anomaly itself and thereby attempt to model this behavior near Saturn encounter along with the near-constant nature of the anomalous `acceleration' long after for Pioneer 11. Furthermore, we look to model the near-constant nature of the anomalous `acceleration' throughout the entire data set analysed for Pioneer 10. We assume throughout our discussion that, in future work when fitting to the extended set of DSN Doppler data, the thermal contribution to the anomaly can be approximately modeled according to a weighted mean of the original analysis of Ref.~\refcite{Anderson} with the analysis of Ref.~\refcite{ThermalPioneer2012}, giving almost $\sim 12$\% of the full observed unmodeled `acceleration' value, reported in Ref.~\refcite{Anderson} to be $a_p = (8.74 \pm 1.33)\times 10^{-13}$ km s$^{-2}$. We will discuss this assumption about the thermal contribution to the anomaly in more detail in our conclusions. To potentially accomplish this task of modeling the remaining nearly $8.0\times10^{-13}$ km s$^{-2}$ of the anomalous `acceleration,' we employ a recently proposed cosmological model termed ``the theory of inertial centers"\cite{Feldman} and use the analysis of Refs.~\refcite{Kopeikin} and \refcite{KopeikinEPJ} as inspiration for our reasoning. However, we must emphasize that we are {\it not} employing the results of Refs.~\refcite{Kopeikin} and \refcite{KopeikinEPJ}. While there have been other examinations of the possibility that the Pioneer anomaly can be linked to a cosmological origin such as the analyses in Refs.~\citen{Mizony2005, Lachieze-Rey2007} and \refcite{Iorio2014}, we will not employ any of these results either given the assumptions of this newly proposed theory. Instead, with an understanding of the fundamental assumptions taken by the DSN, we derive in this article our own results relevant to the predicted DSN frequency residuals according to the theory of inertial centers. Taking input data from the {\footnotesize HORIZONS} system of Jet Propulsion Laboratory (JPL) solely for calculating approximate values of our predicted residual term (i.e. our analysis is a rough `first-order' approximation prior to proper fitting to DSN frequency data) and adopting a seemingly arbitrary convention due to a mathematical ambiguity to relate physical clock-rates in this model to DSN-assumed clock-rates, we discover that this theory is able to replicate the near-constant and blueshifted nature of the anomalous `acceleration' for the latter interval of cruise phase out of the Solar System for both Pioneer 10 and Pioneer 11 as well as a sudden onset in anomalous `acceleration' after Saturn encounter by Pioneer 11. Thus, we tentatively suggest that the predictions of this cosmological model could provide an alternative to the thermal explanation of the Pioneer anomaly. However, a truly convincing explanation for the adoption of our particular convention for relating clock-rates is paramount to our understanding. Unfortunately, this resolution has yet to be determined. Still, given the ability with this convention to predict an onset for Pioneer 11 after Saturn encounter, we suggest the continuation of related work to further explore the consequences of these predictions in the form of potential observables for ongoing experiments in our solar system.

This article is organized in the following manner. We begin with a brief review of the concepts employed in Ref.~\refcite{Kopeikin} and thoroughly detailed in Ref.~\refcite{Moyer}. These are vital to our understanding of how the DSN calculates frequency residuals. We then review the principles presented in Ref.~\refcite{Feldman} most relevant for our purposes, subsequently deriving an additional clock-drift term according to the theory of inertial centers for frequency residuals calculated by the DSN using the geodesic solutions of \ref{GeodesicSolutions}. As mentioned previously, we employ a particular convention for the relation between physical clock-rates in this theory and DSN-assumed clock-rates in our derivation. We compare and contrast the ``clock acceleration" (see section~V.B of Ref.~\refcite{Anderson}) predicted here with that of Ref.~\refcite{Kopeikin}, emphasizing an additional weighting factor in the theory of inertial centers' prediction dependent upon the direction of propagation of our transmitted photons relative to the direction of the center of our galaxy. This additional weighting factor, unique to the arguments presented in this article, produces a sudden onset of the predicted anomalous behavior after Saturn encounter by Pioneer 11. We provide concluding remarks and mention relevant future work.

\section{Results and Discussion}

As alluded to in our introduction, the analysis in this section is very much inspired by the work of Ref.~\refcite{Kopeikin}, which attempts to provide an explanation for the Pioneer anomaly within the context of our current standard model of cosmology. We offer a brief review to prepare for our own approach with the theory of inertial centers. The key feature pointed out in Ref.~\refcite{Kopeikin} is that current astrometric practice ignores the difference between coordinates suitable for an expanding universe and coordinates appropriate for globally flat space-time in general relativity after ``factoring out" effects from gravitational sources. Presently, we take
\begin{equation} \label{lightray}
c(t_2 - t_1) = r_{21},
\end{equation}
where $r_{21} = |{\mathbf r_2} -{\mathbf r_1}|$ is the Euclidean coordinate distance between the point of emission and the point of reception of light. The DSN assumes this is equal to the proper distance $d_{21}$ in our empty vacuum scenario. Furthermore, $c$ is the DSN constant for the speed of light in vacuum and $t_2 - t_1$ is the change in coordinate time between emission and reception.

However, in the Friedmann-Lema\^{i}tre-Robertson-Walker (FLRW) metric,\cite{Friedmann} one must take into account the rate of assumed expansion of the universe with the Hubble constant $H_0$ \cite{Hubblelaw, Riess} to linear order for our current experimental precision {\it with photons}. To account for this expansion, Ref.~\refcite{Kopeikin} suggests that the above equation must be replaced by
\begin{equation}
c(\lambda_2 - \lambda_1) = r_{21},
\end{equation}
where
\begin{equation}
\lambda_{i} = t_i + \frac{1}{2} H_0 t_i^2,
\end{equation}
analogous to the phenomenological ``quadratic in time model" of Section~XI.E in Ref.~\refcite{Anderson}. The Newtonian approximation is assumed so that each observer is taken to be nearly stationary relative to an observer abiding by barycentric coordinate time (TCB) such that $t = \tau$, where $\tau$ is the proper time of the observer measuring frequency shifts. In other words, it is suggested that to continue using a light-time relation of the form in (\ref{lightray}) for empty space while still accounting for an expanding universe when modeling away from gravitational sources, one must take the light-ray coordinates to $\mathcal{O}(H_0)$ to be $\{\lambda, {\mathbf r} \}$ and {\it not} $\{t, {\mathbf r} \}$. Ref.~\refcite{Kopeikin} goes on to find that the two-way fractional frequency shift calculated by the DSN for one round-trip will have an additional term
\begin{equation}
\frac{\nu_3}{\nu_1} = 1- 2[\beta_{21} - H_0(t_2 - t_1)],
\end{equation}
which would give the desired blueshift to possibly explain the anomalous acceleration in frequency shift seen with the Pioneer spacecrafts once we integrate over multiple round-trip cycles as the clock acceleration $a_t$ quoted in Ref.~\refcite{Anderson} using the formula
\begin{equation}
\frac{\nu_3}{\nu_1}\bigg|_{\mathrm{observed}} - \frac{\nu_3}{\nu_1}\bigg|_{\mathrm{model}} =  2a_t (t_2 - t_1),
\end{equation}
is similar in value to that of the Hubble constant. $\beta_{21}$ is the relative velocity between the spacecraft and observer on Earth along the direction of motion of the light signal given by $\hat{\mathbf k}_{21} \cdot ({\mathbf v}_2 - {\mathbf v}_1)/c$ and $\hat{\mathbf k}_{21}$ is the unit vector directed toward the spacecraft from the observer on Earth.

With the above summary as a preface, we apply this understanding of current practice to the predictions of the theory of inertial centers. This theory suggests that in order to account for the observed effects which our current standard cosmological model (i.e. $\Lambda$-CDM) attributes to the theoretical concepts termed `dark energy' and `dark matter,' one need not assume the existence of these concepts, but instead {\it one may reproduce this behavior by redefining what we consider to be inertial motion as well as what constitutes an inertial frame of reference in globally flat space-time}. However, locally within these redefined inertial systems, this theory was shown to abide by the axioms of general relativity.\cite{Feldman} The conclusions presented in Ref.~\refcite{Feldman} and differences with respect to general relativity result from the form of the affine parameter assumed in this theory for massive geodesics,
\begin{equation}\label{eq:affine}
\chi = \sqrt{\Lambda} \int r(\tau) d\tau,
\end{equation}
where $\sqrt{\Lambda}$ is taken to be a universal constant and assumed to be our measured value of the Hubble constant. Furthermore, $r(\tau)$ is the observer's physical distance to the center of one of these newly defined inertial systems and $\tau$ represents the physical clock-rate of the observer in this particular inertial system. Meaning, the physically observable elapsed time as measured by a clock carried along a given curve, denoted as proper time $\tau$, is {\it not} the affine parameter for the theory of inertial centers and thus is not invariant in this theory. Therefore, observable clock-rates depend upon both spatial position in a particular inertial frame as well as in which inertial frame the observer is observing. In contrast, the affine parameter for massive geodesics in general relativity takes the form $\chi = c \int d\tau$, where instead $c$ is taken to be a universal constant as the speed of light in globally flat space-time (see ch.~4.2 and 4.3 of Ref.~\refcite{Wald_GR}).

To be explicit with regard to inertial motion, according to the theory of inertial centers, a massive object follows a geodesic path {\it affinely parametrized by} $\chi$ (and not by $\tau$ as in general relativity) when subjected to no net external forces with equations of motion given by
\begin{equation} \label{eq: eqofmotion}
0 = U^a \nabla_a U^b,
\end{equation}
where $U^a$ is the tangent vector to the geodesic curve denoted $x^{\alpha}(\chi)$ in a particular coordinate system (i.e. $U^{\alpha} = dx^{\alpha}/d\chi$). Notice that we have employed abstract index notation and will continue to do so throughout this article (see ch.~2.4 of Ref.~\refcite{Wald_GR}). Expanding (\ref{eq: eqofmotion}) in a particular coordinate system, one finds that the equations of motion for our massive object reduce to
\begin{equation}
0 = \frac{d^2 x^{\alpha}}{d \chi^2} + \sum_{\mu, \nu} \Gamma^{\alpha}_{\mu \nu} \frac{dx^{\mu}}{d\chi} \frac{dx^{\nu}}{d\chi},
\end{equation}
where $\Gamma^{\alpha}_{\mu \nu}$ are the components of the affine connection for the coordinate chart that we are employing to describe the geodesic path. However, to describe this path in terms of physical observables (i.e. proper time $\tau$), one must use the conjectured relation between $\chi$ and proper time in the particular inertial system within which the massive object moves. We have
\begin{equation}\label{geo}
0 = \frac{d^2 \tau}{d\chi^2} \frac{dx^{\alpha}}{d\tau} + \bigg(\frac{d\tau}{d\chi}\bigg)^2 \bigg[\frac{d^2 x^{\alpha}}{d \tau^2} + \sum_{\mu, \nu} \Gamma^{\alpha}_{\mu \nu} \frac{dx^{\mu}}{d\tau} \frac{dx^{\nu}}{d\tau}\bigg],
\end{equation}
where $d\chi/d\tau = \sqrt{\Lambda} \cdot r(\tau)$. Additionally, if the bracketed term in (\ref{geo}) were to equal zero, it is evident that we would have the geodesic equations of motion according to general relativity,\cite{Wald_GR} and thus (\ref{geo}) shows the clear physical difference between the geodesics of the theory of inertial centers and those of general relativity (i.e. in general, $d^2\tau/d\chi^2 \cdot dx^{\alpha}/d\tau \neq 0$; the general relativistic limit occurs for $r(\tau) =$ constant). In terms of the proper velocity of the massive object $u^a$ with components $u^{\alpha} = dx^{\alpha}/d\tau$, our geodesic equation implies
\begin{equation}
u^a \nabla_a u^b = \frac{1}{r} \frac{dr}{d\tau} u^b,
\end{equation}
in accordance with eqn.~(3.3.2) of Ref.~\refcite{Wald_GR}. 

To describe one of these newly proposed inertial frames of reference, it appears ``most natural" to employ the radial Rindler chart
\begin{eqnarray}\label{metric}
ds^2 = -\Lambda r^2 dt^2 + dr^2 + r^2 \cosh^2(\sqrt{\Lambda}t)(d\theta^2 + \sin^2{\theta}d\phi^2),
\end{eqnarray}
where $\sqrt{\Lambda} = H_0$ is taken to be the Hubble constant and $ds^2 = -d\chi^2 = -\Lambda r^2 d\tau^2$ for massive geodesics, as coordinate time $t$ in this chart corresponds to the physical time elapsed on the clock of a stationary observer according to the assumptions of this theory (i.e. inertial time). Notice that observable clock-rates in this theory differ from those of general relativity due to the form of the affine parameter and reflected in the fact that, in globally flat space-time, they are dependent upon {\it both} the relative velocity between observers as well as relative spatial position with respect to the center of the inertial frame. Ref.~\refcite{Feldman} terms the center of each new inertial frame an ``inertial center" (in this chart, $r=0$ is taken to be the location of our inertial center).

In the following, we use the term ``stationary observer" to refer to any observer that is stationary relative to the center of the inertial frame in question as defined according to the theory of inertial centers. Furthermore, Ref.~\refcite{Feldman} takes the Milky Way itself to constitute one of these new inertial frames with the center of the Milky Way in the direction of Sagittarius A* constituting the center of the inertial reference frame relevant for observers in our solar system. In other words, it is assumed in this model that at the center of our galaxy there exists an inertial center and {\it not} a supermassive black hole (for recent tension with our current understanding of black hole formation, see Refs.~\refcite{MersiniA} and \refcite{MersiniB}). However, as Ref.~\refcite{Kopeikin} mentions in his work, the adoption for the time coordinate in current programs assumes the Minkowski coordinate parametrization of space-time to be accurate for completely empty space, yet this is also not the case for the theory of inertial centers. Meaning, the DSN assumes that Minkowski time is consistent with the progression of a physical clock in deep-space far from gravitational sources, which in fact is {\it not} the case when operating under the assumptions of this theory. Therefore, we should formulate calculated frequency shifts in terms of the coordinate time assumed by the DSN to determine possible residuals. We emphasize the word ``calculating" because this additional `frequency shift' is not actually a physical shift in the frequency of our two-way electromagnetic signals according to this model (see the end of \ref{gendopplershift}). Instead, it is conjectured that the DSN's definition of time resulting from the assumed validity of (\ref{lightray}) in empty space is inconsistent with the manner in which physical clocks progress. This approach is in contrast to the naive procedure outlined in ``Application to a local gravitational system" in Ref.~\refcite{Feldman} where the Pioneer anomaly is suggested to be the result of a physical frequency shift.

\subsection{Stationary observer treatment}

For the general case of examining the two-way frequency residuals as calculated by the DSN from light signals sent between a spacecraft in our solar system and our tracking stations here on Earth, we must analyse the range-rate calculation employed in Ref.~\refcite{Moyer} with the understanding that the DSN's definition of time does {\it not} coincide with physical clock-time according to the theory of inertial centers. Thus, we look to relate clock-rates by deriving the differential change in DSN-assumed coordinate time with respect to physical time in this theory. Our approach to accomplishing this task relies upon relating the light-time expected by the DSN with the light-time expected by the theory of inertial centers as this quantity, in essence, defines changes in time for a fixed proper distance traversed (i.e. a light clock). We refer to the geodesic solutions for a light signal traveling within the Milky Way inertial system with an inertial center assumed to be located in the direction of Sagittarius A*. From (\ref{radialmotion}) and (\ref{masslesslightspeed}) in \ref{GeodesicSolutions}, we have for the radial solution of the geodesic path of a massless particle
\begin{equation}
r(t) = r_0 \bigg[\frac{v_0^r}{c_0}\sinh(\sqrt{\Lambda}t) + \sqrt{1 + \bigg(\frac{v_0^r}{c_0}\bigg)^2 \sinh^2(\sqrt{\Lambda}t) } \bigg],
\end{equation}
where $v_0^r = dr/dt|_{t=0}$ is the initial radial velocity of our particle at $t=0$, $r_0 = r(t=0)$ is the initial distance to our inertial center, and $c_0 = \sqrt{\Lambda}r_0$. Additionally, $v_0 = c_0 = \sqrt{\sum_{i} v^i v_i|_{t=0}} $ is the initial speed of our massless particle where this sum occurs over a constant $t$ spatial hypersurface of our metric (\ref{metric}) with induced Riemannian metric components $h_{ij}$. Solving for the elapsed coordinate time $t$ for a light signal traveling between $r_0$ and $r$
\begin{equation}
t = \frac{1}{\sqrt{\Lambda}} \sinh^{-1} \bigg[\frac{c_0}{v_0^r} \sinh\bigg(\ln (r/r_0) \bigg) \bigg],
\end{equation}
but $v_0^r = c_0 (\hat{\mathbf k}_0 \mathbf{\cdot} \hat{\mathbf r}_0)$, where $\hat{\mathbf k}_0$ is the initial direction of propagation of our light signal and $\hat{\mathbf r}_0$ is the initial radial unit vector for the emitter of said signal within our inertial system. Then,
\begin{equation}
t_{\pm} = \frac{1}{\sqrt{\Lambda}} \sinh^{-1} \bigg[\frac{1}{|\hat{\mathbf k}_0 \mathbf{\cdot} \hat{\mathbf r}_0|} \sinh\bigg(\pm \ln (r/r_0) \bigg) \bigg],
\end{equation}
where $t_{+}$ is the elapsed coordinate time for a signal traveling away from our inertial center and $t_{-}$ is the elapsed coordinate time for a signal traveling toward our inertial center. For a two-way signal sent between nearly stationary observers labeled $1$ and $2$ within the galactic inertial system, notice that
\begin{eqnarray}
t_{\mathrm{uplink}} = \frac{1}{\sqrt{\Lambda}} \sinh^{-1} \bigg[\frac{1}{|\hat{\mathbf k}_1 \mathbf{\cdot} \hat{\mathbf r}_1|} \sinh\bigg(\pm \ln (r_2/r_1) \bigg) \bigg]
\\ t_{\mathrm{downlink}} = \frac{1}{\sqrt{\Lambda}} \sinh^{-1} \bigg[\frac{1}{|\hat{\mathbf k}_2 \mathbf{\cdot} \hat{\mathbf r}_2|} \sinh\bigg(\mp \ln (r_1/r_2) \bigg) \bigg].
\end{eqnarray}
For signals sent within the Solar System, we will make the approximation $\hat{\mathbf k}_1|_{\mathrm{uplink}} = \hat{\mathbf k}_2|_{\mathrm{uplink}}= -\hat{\mathbf k}_2|_{\mathrm{downlink}} $ and assume our galactic radial unit vector is approximately the same everywhere within our solar system (i.e. $\hat{\mathbf r}_1 = \hat{\mathbf r}_2 = \hat{\mathbf r}$). Thus, $t_{\mathrm{uplink}} = t_{\mathrm{downlink}}$. Then, the elapsed coordinate time for a light signal sent one-way between observers located at two radial distances $r_b$ and $r_s$ is given by
\begin{equation}\label{genlighttime}
t = \frac{1}{\sqrt{\Lambda}} \sinh^{-1} \bigg[\frac{1}{|\hat{\mathbf k}_s \mathbf{\cdot} \hat{\mathbf r}|} \sinh\bigg( \ln (r_b/r_s) \bigg) \bigg],
\end{equation}
where $r_b > r_s$ (i.e. $r_b = r_2$, $r_s = r_1$ for $r_2>r_1$ and $r_b = r_1$, $r_s = r_2$ for $r_2 < r_1$).

The DSN assumes the elapsed coordinate time $T_{\mathrm{DSN}}$ for light traveling a total {\it physical} distance $d_{21}$ (i.e. proper distance) is given by\cite{Moyer}
\begin{equation}\label{Minktime}
T_{\mathrm{DSN}} = \frac{d_{21}}{c_{\mathrm{DSN}}},
\end{equation}
where $c_{\mathrm{DSN}}$ is the DSN-assumed value for the speed of light in empty vacuum. Yet, given our assumptions of the near-constancy of the direction of our galactic radial unit vector within the Solar System, we find that the DSN assumes the physical radial distance traversed by a one-way light signal in the galactic reference frame should be given in terms of DSN coordinate time by
\begin{equation}\label{eq:GeneralDSNTime}
|\Delta r| \approx |\hat{\mathbf k}_s \mathbf{\cdot} \hat{\mathbf r}| c_{\mathrm{DSN}}T_{\mathrm{DSN}}.
\end{equation}
We will use this as our analogue to the defining relation for DSN light-time (\ref{Minktime}) in the rest of our analysis. As well, we take our proper radial length in the galactic frame according to the theory of inertial centers to be $|\Delta r| = |r_2 - r_1| = r_b - r_s$.

In looking for an expression relating DSN-assumed clock-rates to physical clock-rates in the theory of inertial centers (i.e. $dT_{\mathrm{DSN}}/dt$), we come upon the unfortunate consequence that our light-time in this model relies upon {\it fractional differences in distance}, whereas the DSN definition of light-time relies upon the actual difference in coordinate values. Thus, the question becomes: where do we place the dependence on $T_{\mathrm{DSN}}$ in our fractional expression $r_b/r_s$ given that the DSN defines time based off of $r_b - r_s$? In other words, expressing the fractional difference in terms of $|\Delta r| = r_b - r_s$, one has
\begin{equation}
\frac{r_b}{r_s} = 1+ \frac{|\Delta r|}{r_s} = \frac{r_b}{r_b - |\Delta r|}.
\end{equation}
But the remaining presence of radial coordinate values $r_b$ or $r_s$ in this fractional difference expression, which we will eventually plug back into (\ref{genlighttime}), inherently forces us to make a choice as to whether or not $r_b$ or $r_s$ has dependence upon $T_{\mathrm{DSN}}$ as we are looking for the differential change in range with respect to time elapsed on the physical clocks of this model. We focus on two possibilities:
\begin{equation}
r_b = r_b(T_{\mathrm{DSN}}) \indent r_s = \textrm{constant},
\end{equation}
or
\begin{equation}
r_b = \textrm{constant} \indent r_s = r_s(T_{\mathrm{DSN}}),
\end{equation}
such that (\ref{eq:GeneralDSNTime}) reproduces a DSN-assumed null geodesic of the form
\begin{equation}\label{former}
r_b(T_{\mathrm{DSN}}) = |\hat{\mathbf {k}}_s \mathbf{\cdot} \hat{\mathbf {r}}|c_{\mathrm{DSN}}T_{\mathrm{DSN}} + r_s,
\end{equation}
for the former case or
\begin{equation}\label{latter}
r_s(T_{\mathrm{DSN}}) = -|\hat{\mathbf {k}}_s \mathbf{\cdot} \hat{\mathbf {r}}|c_{\mathrm{DSN}}T_{\mathrm{DSN}} + r_b,
\end{equation}
for the latter. Again, we've assumed $|\hat{\mathbf k}_s \mathbf{\cdot} \hat{\mathbf r}| = |\hat{\mathbf k}_b \mathbf{\cdot} \hat{\mathbf r}|$ for the one-way signal. Note that these DSN-assumed null geodesic equations are valid only for our assumptions of an electromagnetic signal traveling within a small region of the galaxy.

We seek to express our fractional difference in radial positions in terms of only assumed constants and $T_{\mathrm{DSN}}$. For the former case, one has in (\ref{genlighttime}) the expression
\begin{eqnarray}
\ln \bigg( \frac{r_b}{r_s} \bigg) = \ln \bigg(1+\frac{| \Delta r|}{r_s} \bigg) \nonumber
\\ = \ln \bigg(1+\frac{|\hat{\mathbf k}_s \mathbf{\cdot} \hat{\mathbf r}| c_{\mathrm{DSN}}T_{\mathrm{DSN}}}{r_s} \bigg),
\end{eqnarray}
where we've used defining relation (\ref{eq:GeneralDSNTime}). If one were to adopt $T_{\mathrm{DSN}}$ as a time coordinate covering our entire inertial system, he/she would find that this expression is valid for all values of $0<T_{\mathrm{DSN}}<\infty$. However, for the latter case, one finds
\begin{eqnarray}
\ln \bigg( \frac{r_b}{r_s} \bigg) = \ln \bigg(\frac{r_b}{r_b - |\Delta r|} \bigg) \nonumber
\\ = \ln \bigg(\frac{r_b}{r_b - |\hat{\mathbf k}_s \mathbf{\cdot} \hat{\mathbf r}| c_{\mathrm{DSN}}T_{\mathrm{DSN}}}  \bigg),
\end{eqnarray}
where this expression restricts our $T_{\mathrm{DSN}}$ `independent' parameter values such that $0<T_{\mathrm{DSN}} < r_b/|\hat{\mathbf k}_s \mathbf{\cdot} \hat{\mathbf r}|c_{\mathrm{DSN}}$. In reality, of course, $T_{\mathrm{DSN}}$ is not an independent parameter as its measured value depends upon the proper distance between our two observers. Nevertheless, referring to (\ref{eq:GeneralDSNTime}), one finds, as of course we should expect, that this restriction on $T_{\mathrm{DSN}}$ is consistent with our expectations for the DSN-assumed light-time since we must have $r_s > 0$ for our current analysis. Still, in this instance, it does not seem that one could use $T_{\mathrm{DSN}}$ as a global `time coordinate' for a DSN parametrization of empty space-time in the inertial system given the restrictions imposed by the theory of inertial centers. While it seems rather arbitrary to suggest that this has any physical relevance on our interpretation and predictions for DSN observations as  both choices for $T_{\mathrm{DSN}}$ are consistent with DSN-assumed light-time conditions, we mention the idea for discussion. Do the restrictions on the allowed values of $T_{\mathrm{DSN}}$ for the latter case point to the suggestion that the DSN time coordinate in this scenario is unphysical and thus the only physical solution lies with the convention $r_b = r_b(T_{\mathrm{DSN}})$? Given these arguments, we'll initially approach this problem of determining DSN residuals according to the theory of inertial centers by adopting the former convention of $r_b = r_b(T_{\mathrm{DSN}})$, where $T_{\mathrm{DSN}}$ as an `independent' parameter can take all values greater than zero in its relationship with $t$.

Manipulating (\ref{genlighttime}) to isolate our fractional difference factor, one finds
\begin{equation}
\frac{r_b}{r_s} = {\mathrm e}^{\sinh^{-1}[|\hat{\mathbf k}_s \mathbf{\cdot} \hat{\mathbf r} | \sinh(\sqrt{\Lambda}t)]}.
\end{equation}
But expressing in terms of $T_{\mathrm{DSN}}$, we have for the former case the relation
\begin{equation}\label{eq:DSNTime}
T_{\mathrm{DSN}}= \frac{r_s}{|\hat{\mathbf k}_s \mathbf{\cdot} \hat{\mathbf r}| c_{\mathrm{DSN}}} \bigg[ {\mathrm e}^{\sinh^{-1}[|\hat{\mathbf k}_s \mathbf{\cdot} \hat{\mathbf r} | \sinh(\sqrt{\Lambda}t)]} -1 \bigg],
\end{equation}
where we again emphasize that the smaller radial coordinate value, $r_s$, is taken to be a constant. Proceeding to the relevant observable, we find that the theory of inertial centers prediction for the differential change in DSN-expected light-time with respect to the physical clock of an observer stationary relative to the center of the Milky Way is given by
\begin{equation}
\frac{dT_{\mathrm{DSN}}}{dt} = \frac{\sqrt{\Lambda}r_s}{c_{\mathrm{DSN}}}\frac{{\mathrm e}^{\sinh^{-1}[|\hat{\mathbf k}_s \mathbf{\cdot} \hat{\mathbf r} | \sinh(\sqrt{\Lambda}t)]}\cosh(\sqrt{\Lambda}t)}{\sqrt{1+(\hat{\mathbf k}_s \mathbf{\cdot} \hat{\mathbf r})^2 \sinh^2(\sqrt{\Lambda}t)}}.
\end{equation}
Expanding to first-order in $\sqrt{\Lambda}t$, we find
\begin{equation}\label{drift1}
\frac{dT_{\mathrm{DSN}}}{dt} = \frac{\sqrt{\Lambda}r_s}{c_{\mathrm{DSN}}}\bigg[1+ |\hat{\mathbf k}_s \mathbf{\cdot} \hat{\mathbf r}| \sqrt{\Lambda}t\bigg]  + \mathcal{O}(\Lambda t^2).
\end{equation}
However, the frequency shift {\it calculated by the DSN} is taken to be the ratio of differential proper times, where it is assumed that the total light-time is given by $T_{\mathrm{DSN}}$.\cite{Moyer} Thus, we take the time coordinate employed by the DSN to be $T_{\mathrm{DSN}}$ and {\it not} $t$ when calculating range-rate. Meaning, in this model, the two-way DSN fractional frequency shift $[\nu_3/\nu_1]_{\mathrm{DSN}}$ does not represent an accurate calculation of the fractional shift in the frequency observable itself (i.e. $\nu_3/\nu_1 \neq [\nu_3/\nu_1]_{\mathrm{DSN}}$) due to the fact that the DSN's definition of time drifts away from actual physical time according to the theory of inertial centers with our first-order relation modeling this drift given by (\ref{drift1}).

Then, the two-way DSN fractional frequency shift between stationary observers ($dt/d\tau|_i = 1$ and $dT_{\mathrm{DSN}}|_i/dT_{\mathrm{DSN}}|_j = 1$ for $i,j = 1,2,3$) is predicted by the theory of inertial centers to be
\begin{eqnarray}
\bigg[\frac{\nu_3}{\nu_1}\bigg]_{\mathrm{DSN}} = \bigg[\frac{dT_{\mathrm{DSN}}}{dt}\bigg|_{2} \bigg(\frac{dT_{\mathrm{DSN}}}{dt}\bigg|_{1}\bigg)^{-1}\bigg]\bigg|_{\mathrm{uplink}} \cdot \bigg[\frac{dT_{\mathrm{DSN}}}{dt}\bigg|_{3} \bigg(\frac{dT_{\mathrm{DSN}}}{dt}\bigg|_{2}\bigg)^{-1}\bigg]\bigg|_{\mathrm{downlink}} \nonumber
\\ = \frac{1+ |\hat{\mathbf k}_2 \mathbf{\cdot} \hat{\mathbf r}| \sqrt{\Lambda}t_2}{1+ |\hat{\mathbf k}_1 \mathbf{\cdot} \hat{\mathbf r}| \sqrt{\Lambda}t_1}\cdot \frac{1+ |\hat{\mathbf k}_3 \mathbf{\cdot} \hat{\mathbf r}| \sqrt{\Lambda}t_3}{1+ |\hat{\mathbf k}_2 \mathbf{\cdot} \hat{\mathbf r}| \sqrt{\Lambda}t_2} + \mathcal{O}(\Lambda t^2) \nonumber
\\ = 1+ |\hat{\mathbf k} \mathbf{\cdot} \hat{\mathbf r}| \sqrt{\Lambda}(t_3 - t_1) + \mathcal{O}(\Lambda t^2).
\end{eqnarray}
assuming $\hat{\mathbf k}_1 = \hat{\mathbf k}_2|_{\mathrm{uplink}} = - \hat{\mathbf k}_2|_{\mathrm{downlink}} = - \hat{\mathbf k}_3 = \hat{\mathbf k}$. However, as we found earlier, $t_{\mathrm{uplink}} = t_{\mathrm{downlink}}$ for our assumptions in this small region of our galaxy, where $t_{\mathrm{uplink}} = t_2 - t_1$ and $t_{\mathrm{downlink}} = t_3 - t_2$. We have
\begin{equation} \label{2wayDSNshift}
\bigg[\frac{\nu_3}{\nu_1}\bigg]_{\mathrm{DSN}} = 1+ 2|\hat{\mathbf k} \mathbf{\cdot} \hat{\mathbf r}| \sqrt{\Lambda}(t_2 - t_1) + \mathcal{O}(\Lambda t^2),
\end{equation}
where $t_2 - t_1$ is the one-way light-time measured by our stationary physical clock. One finds for the two-way frequency shift as calculated by the DSN an additional drift term with clock acceleration given by
\begin{equation}\label{clockaccel}
a_t = \sqrt{\Lambda} |\hat{\mathbf k} \mathbf{\cdot} \hat{\mathbf r}|.
\end{equation}
Furthermore, with the adopted convention of $r_b = r_b(T_{\mathrm{DSN}})$, our clock-drift term produces a {\it blueshifted} residual independent of whether or not the signal initially travels away from or toward the galactic center. Notice that one obtains the result of Ref.~\refcite{Kopeikin} for photons traveling either directly toward or away from the galactic center as $\sqrt{\Lambda}$ is taken to have the value of the Hubble constant in the theory of inertial centers. Since Pioneer 10 and Pioneer 11 traveled nearly directly away from and toward the center of our galaxy, respectively, on their way out of the Solar System\cite{TuryshevToth} (i.e. $|\hat{\mathbf k} \mathbf{\cdot} \hat{\mathbf r}| \approx 1$), these results appear at first glance to be consistent with the late-time observations from Ref.~\refcite{Anderson} of a blueshifted clock-drift residual for both Pioneer 10 and Pioneer 11 as the observed $a_t$ was similar in value to that of the Hubble constant.

Nevertheless, the key to comparing these predictions with actual DSN residuals lies in calculating the additional drift term from (\ref{2wayDSNshift}). In order to proceed, we require the complete history of the path of the spacecraft relative to Earth's path over the course of our observations to calculate the direction of propagation relative to the center of the Milky Way of each light signal sent between our stations on Earth and the spacecraft. Still, the above equation is very much an approximation because we assumed both the spacecraft and Earth observer are stationary relative to one another and chose to neglect gravitational effects. However, one can think of this analysis as ``factoring out" the Doppler shift due to the relative velocity of our observers as well as any potential frequency shift due to gravitational sources (see \ref{gendopplershift} for a more general derivation taking into account the Doppler effect). Meaning, additional terms predicted by this theory with respect to the DSN's expectations in a full relativistic derivation will be suppressed by the multiplication of $\sqrt{\Lambda} t$ and thus can likely be disregarded to first-order in $\sqrt{\Lambda}t, (v/c),$ and $GM/c^2d_M$ where $d_M$ is the distance to the center of a massive object in our solar system. Furthermore, the weighting factor $|\hat{\mathbf k} \mathbf{\cdot} \hat{\mathbf r}|$ in our clock acceleration reflects the amount that the DSN's definition of time coincides with physical time according to the theory of inertial centers (see (\ref{drift1})), as the difference in light-time between that assumed by the DSN in the Solar System and what physically occurs in the theory of inertial centers is smallest when both observers are located the same distance away from an inertial center (i.e. $|\hat{\mathbf k} \mathbf{\cdot} \hat{\mathbf r}| = 0$) and largest when the photon travels radially between observers located at two different distances from the center of our galaxy (i.e. $|\hat{\mathbf k} \mathbf{\cdot} \hat{\mathbf r}| = 1$).

While examining alternatives to the thermal explanation for the Pioneer anomaly, one may ask and rightly so: if the initial results in our DSN calculations of two-way frequency shifts for Pioneers 10 and 11 are nearly the same as those produced by Ref.~\refcite{Kopeikin}, how can we tell whether or not the Pioneer anomaly points to a consequence of the assumptions of the theory of inertial centers or to a consequence of the analysis of Ref.~\refcite{Kopeikin}? Unfortunately, due to the paths of both Pioneer spacecrafts almost directly toward and away from Sagittarius A* on their way out of the Solar System, it would be difficult to tell the difference between the predictions of both models with the precision of the available data for the average late-time behavior of the anomaly. However, {\it the results of Ref.~\refcite{Kopeikin} are based upon assumptions of isotropy}. Thus, they should hold for spacecrafts moving in any particular direction in outer space far from the Sun in the galactic frame. The same is {\it not} true for the theory of inertial centers as one must consider the paths of these spacecrafts {\it relative to the center of the Milky Way}. Therefore, if we were to send a spacecraft out moving tangentially to the direction of the center of the Milky Way, we would obtain a different result while working under the assumptions of the theory of inertial centers versus those of Ref.~\refcite{Kopeikin}. For context in the rest of our analysis, the predicted residuals calculated by the DSN should vary according to the trajectory of the spacecraft relative to our DSN stations on Earth if one models with the theory of inertial centers. Thus, a gravitational assist that significantly alters the trajectory of a spacecraft, such as that of Pioneer 11 by Saturn, could potentially ``turn on" the anomaly according to this theory, whereas this type of ``turn on" would not seem to be possible according to the analysis of Ref.~\refcite{Kopeikin}.

Before proceeding to a more comprehensive analysis of our predictions for several spacecrafts, we must return to the latter scenario of our convention choices with $r_s = r_s(T_{\mathrm{DSN}})$ for our dependence on DSN-defined time. We recall the restrictions on this `time coordinate'
\begin{equation}
0 < T_{\mathrm{DSN}} <\frac{r_b}{|\hat{\mathbf k}_s \mathbf{\cdot} \hat{\mathbf r}|c_{\mathrm{DSN}}},
\end{equation}
which led us to the suggestion that we can possibly consider this scenario to be unphysical as this candidate for a DSN time coordinate cannot cover all relevant physical values in a relation between $T_{\mathrm{DSN}}$ and $t$. Still, as mentioned in our abstract, we are not completely convinced by this argument. Working through a similar analysis as that employed for the former case, one finds to first-order in $\sqrt{\Lambda}t$
\begin{equation}
\frac{dT_{\mathrm{DSN}}}{dt} = \frac{\sqrt{\Lambda}r_b}{c_{\mathrm{DSN}}}\bigg[1- |\hat{\mathbf k}_s \mathbf{\cdot} \hat{\mathbf r}| \sqrt{\Lambda}t\bigg]  + \mathcal{O}(\Lambda t^2),
\end{equation}
which would ultimately produce a predicted {\it redshift} in DSN residuals. Therein lies the major issue. The adoption of a particular convention results in a change in sign of our drift term. Yet, we have chosen the former convention mainly because it is consistent with the observations from Ref.~\refcite{Anderson}. Future work will require investigating the validity of our proposed argument that the unphysical nature of the restrictions imposed by the theory of inertial centers rules out the latter scenario.

For emphasis in our next section when speaking of the consistency of these results with our current experimental precision of the ephemerides which tend to be dominated by ranging measurements,\cite{Folknerepm2014, Pitjevaepm2014, Fienga_INPOP10a} we examine how the predictions of the theory of inertial centers affect DSN calculations of range. The important point to take away is that range represents a time-of-flight calculation {\it for a single round-trip} of light signals sent back-and-forth to objects in the Solar System. Contrast this with DSN Doppler observables (i.e. range-rate), which represent a change in Doppler cycle count over a predetermined amount of time where the cycle count is calculated from an integration of the cycles of frequency accumulated over the count time, thus producing a {\it compounding} effect when analyzed over long data intervals (see ch.~13.3 of Ref.~\refcite{Moyer}). Theoretically, one can approximate the manner in which the DSN reproduces this compounding effect by multiplying together multiple fractional frequency ratios as if the signals are bouncing back-and-forth between Earth DSN stations and spacecrafts within the Solar System (i.e. $\nu_n/\nu_1 = (\nu_n/\nu_{n-1})(\nu_{n-1}/\nu_{n-2})\cdots(\nu_2/\nu_1)$). However, this would technically be inaccurate when comparing with the actual implementation detailed in Ref.~\refcite{Moyer}, as remarked upon in section IX.C of Ref.~\refcite{Kopeikin}, because, after a signal's two-way round-trip, the DSN retransmits another signal {\it at the same frequency at which it initially emitted}, $\nu_1$, but time-stamps the point of reception of the previous signal (i.e. $t_3$) and records the difference in frequency between the emitted and received frequency, $\nu_3$ versus $\nu_1$. While one still accumulates phase in a ``compounding" manner with the DSN-employed time-integration process, theoretically it appears simpler to see this effect with a multiplication of multiple fractional frequency ratios. Nevertheless, our point here is to emphasize that ranging measurements do {\it not} have this ``compounding" effect as they only measure a single round-trip for each light signal.

To determine possible additional terms in DSN-range calculations according to the theory of inertial centers, we return to our expression for the DSN-calculated one-way light-time (\ref{eq:DSNTime}) and realize from (\ref{Minktime}) and (\ref{eq:GeneralDSNTime}) that given the manner in which our DSN time coordinate is defined, the actual physical range predicted by the theory of inertial centers will be $\rho = c_{\mathrm{DSN}} T_{\mathrm{DSN}}$. Thus, we have for our range expression when employing our former convention (\ref{former})
\begin{eqnarray}
\rho = c_{\mathrm{DSN}}T_{\mathrm{DSN}} \nonumber
\\ = \frac{r_s}{|\hat{\mathbf k}_s \mathbf{\cdot} \hat{\mathbf r}| } \bigg[ {\mathrm e}^{\sinh^{-1}[|\hat{\mathbf k}_s \mathbf{\cdot} \hat{\mathbf r} | \sinh(\sqrt{\Lambda}t)]} -1 \bigg].
\end{eqnarray}
For light-times within our solar system, we examine terms to second-order in $y=\sqrt{\Lambda}t$ by Taylor expanding about $y = 0$. One has
\begin{eqnarray}
\rho(y=0) = 0
\\ \frac{d\rho}{dy}\bigg|_{y=0} = r_s
\\ \frac{d^2\rho}{dy^2}\bigg|_{y=0} = r_s |\hat{\mathbf k}_s \mathbf{\cdot} \hat{\mathbf r}|,
\end{eqnarray}
which gives
\begin{eqnarray}
\rho = \rho(y=0) + \frac{d\rho}{dy}\bigg|_{y=0} \cdot y +  \frac{1}{2} \frac{d^2\rho}{dy^2}\bigg|_{y=0} \cdot y^2 + \ldots \nonumber
\\ = \sqrt{\Lambda} r_s \bigg[t + \frac{1}{2}|\hat{\mathbf k}_s \mathbf{\cdot} \hat{\mathbf r}| \sqrt{\Lambda}t^2 + \ldots \bigg],
\end{eqnarray}
where $t$ is the actual time-of-flight in this theory. As the DSN {\it expects} range to be given by\cite{Moyer} $\rho_{\mathrm{DSN}} = c_{\mathrm{DSN}} \cdot t$, we'll have as residuals $\Delta \rho_{\mathrm{DSN}}$
\begin{eqnarray}
\Delta \rho_{\mathrm{DSN}} = \rho_{\mathrm{DSN}} - \rho = (c_{\mathrm{DSN}} - \sqrt{\Lambda}r_s)t - \sqrt{\Lambda} r_s \bigg[\frac{1}{2}|\hat{\mathbf k}_s \mathbf{\cdot} \hat{\mathbf r}| \sqrt{\Lambda}t^2 + \ldots \bigg] \nonumber
\\ = \bigg(1- \frac{\sqrt{\Lambda}r_s}{c_{\mathrm{DSN}}} \bigg)\rho_{\mathrm{DSN}} - \frac{1}{2}\sqrt{\Lambda}|\hat{\mathbf k}_s \mathbf{\cdot} \hat{\mathbf r}| \bigg(\frac{\sqrt{\Lambda}r_s}{c_{\mathrm{DSN}}} \bigg) \bigg(\frac{\rho^2_{\mathrm{DSN}}}{c_{\mathrm{DSN}}} \bigg) + \ldots.
\end{eqnarray}

But from ``Reduction to special relativity" in Ref.~\refcite{Feldman} as well as \ref{GeodesicSolutions} and \ref{gendopplershift}, one realizes that $\sqrt{\Lambda}r_s$ actually represents the local speed of light that the observer located a distance $r_s$ from the center of our galaxy measures in this theory. Thus, the value adopted by the DSN for $c_{\rm DSN}$ should correspond to a measurement for the local speed of light by a observer stationary relative to the galactic center and located within our solar system according to this model (in particular, measured at a point in Earth's orbit about the Sun). Taking $c_{\mathrm{DSN}} = \sqrt{\Lambda}r_{\mathrm{DSN}}$, one can express $r_s$ in terms of our DSN observer's distance to the galactic center with
\begin{equation}
r_s = r_{\mathrm{DSN}} + \Delta r,
\end{equation}
where $\Delta r$ is the difference in distance to our galactic center between the locations of our two hypothetical observers $r_s$ and $r_{\mathrm{DSN}}$ (i.e. range in the radial direction in the galactic frame). Notice that if we are thinking in terms of our DSN stations on Earth sending light signals out to our spacecrafts then $r_s/r_{\mathrm{DSN}}$ will only deviate away from one for spacecrafts located {\it closer} to the galactic center and located at large distances away from Earth. One has
\begin{equation}
\Delta \rho_{\mathrm{DSN}} = -\frac{\sqrt{\Lambda}\Delta r}{c_{\mathrm{DSN}}} \rho_{\mathrm{DSN}} - \frac{1}{2}\sqrt{\Lambda}|\hat{\mathbf k}_s \mathbf{\cdot} \hat{\mathbf r}| \bigg(1+ \frac{\sqrt{\Lambda}\Delta r}{c_{\mathrm{DSN}}} \bigg) \bigg(\frac{\rho^2_{\mathrm{DSN}}}{c_{\mathrm{DSN}}} \bigg) + \ldots.
\end{equation}
However, for order of magnitude purposes, we examine the largest possible $\Delta r$ value such that the spacecraft is in exactly the same direction as our galactic center from the Earth observer's perspective. Then for Earth-Mars ranging and Lunar Laser Ranging (LLR) one expects as upper limits $\Delta \rho_{\mathrm{DSN}} \sim .1$ cm and $\Delta \rho_{\mathrm{DSN}} \sim 10^{-7}$ cm, respectively, similar in magnitude to the range results obtained in Kopeikin's analysis (see section IX.B of Ref.~\refcite{Kopeikin}). Nevertheless, our current experimental accuracy for ranging to Mars is of the order of $\sim 10$ m and for LLR $\sim 1$ cm. Therefore, these ranging experiments would be unable to detect the predictions of this theory.

\subsection{Predictions using data from {\footnotesize HORIZONS}}

To facilitate comparison with plots and values from Ref.~\refcite{Anderson}, we formulate all of our predictions for this model in terms of an `acceleration' on our spacecrafts, defining $a_p  \equiv c_{\mathrm{DSN}}a_t= \sqrt{\Lambda}c_{\mathrm{DSN}} |\hat{\mathbf k} \mathbf{\cdot} \hat{\mathbf r}|$, where $c_{\mathrm{DSN}}$ is the DSN-assumed value for the speed of light in vacuum and, as mentioned earlier, we only investigate for the adopted convention of $r_b = r_b(T_{\mathrm{DSN}})$ which predicts a blueshifted clock-drift residual. However, one must keep in mind that this `acceleration' is fictitious in the sense that the spacecraft is {\it not} experiencing an additional acceleration according to the theory of inertial centers. Instead this `acceleration' is a reflection of the difference in clock-rates between physical clocks in the theory of inertial centers and those assumed by the DSN as the clock acceleration $a_t$ is what is truly physically relevant for this model affecting {\it only} the DSN-calculated frequency data. Furthermore, in this theory, one would expect all objects in the Solar System to be moving along a similar geodesic path {\it about the center of our galaxy} and thus, while that geodesic should differ from what one would expect according to general relativity when analyzing in the galactic frame, {\it relative accelerations between objects in the Solar System frame of reference should not differ significantly from the expectations of the DSN as this theory reduces to general relativity within a localized region of the galaxy when objects remain nearly the same distance away from the galactic center} (see ``Reduction to special relativity" and ``Limitations of the study, open questions, and future work" in Ref.~\refcite{Feldman}). Then, one would not expect the positions of objects in the Solar System {\it relative to one another} to differ significantly from DSN expectations when operating under the assumptions of the theory of inertial centers, and thus we do not expect astrometric observations in the ephemerides to be different from general relativistic expectations. As with Ref.~\refcite{Kopeikin}, the only observable differences with our current experimental sensitivity should occur with light, and in particular Doppler measurements due to their ``compounded" nature accentuating that difference, thereby leading to our potential explanation of the Pioneer anomaly. Nevertheless, as shown at the end of our previous section, ranging measurements in the ephemerides will {\it not} be affected according to this theory given our current experimental precision. Since the majority of observations that comprise the ephemerides are ranging and astrometric/optical in nature, it does not appear that terms predicted by this theory would alter the current general relativistic expectations of the ephemerides. Thus, the predictions of this theory should not conflict with our current understanding of the motions of large massive bodies in the Solar System, experimentally verified to a very precise degree.\cite{Iorio_Giudice, Fienga_gravtests, Standish_testing, Standish, Iorio_orbeffects, Iorio_lensethirringPioneer, Iorio_NeptunePioneer, Iorio_velocitydepend, Iorio_Pioneergravorigin, Iorio_JupiterSaturn, Iorio_oort, Page_HowWell, Varieschi2012, Page_MinorPlanets, Page_WeakLimit, Wallin_TestinggravOutersolar, Tangen2007, Folknerepm2014, Pitjevaepm2014, Fienga_INPOP10a, Pitjev_Pitjeva_constraintsdarkmatter, Pitjev_Pitjeva_Reldarkmatter} For our purposes, we approximate the DSN-assumed value for the speed of light in vacuum as $c_{\mathrm{DSN}} = 3.00 \times 10^5$ km s$^{-1}$.

\begin{table}[]
\tbl{Replication of table~I from the original Pioneer anomaly analysis.\cite{Anderson} Determinations of $a_p$ in units of 10$^{-13}$ km s$^{-2}$ from the three time intervals of Pioneer 10 data and from Pioneer 11. JPL analysis defined the time intervals as I (3 January 1987 to 17 July 1990); II (17 July 1990 to 12 July 1992); and III (12 July 1992 to 22 July 1998). Results from various ODP-{\footnotesize SIGMA} and {\footnotesize CHASMP} calculations are listed. ``WLS" signifies a weighted least-squares calculation. ``BSF" signifies a batch-sequential filter calculation. ``With solar corona model" designations refer to methods using the Cassini solar corona model. Lastly, ``Corona, weighting, and F10.7" designation refers to using the Cassini solar corona model with corona data weighting and F10.7 time variation calibration. Errors given are only formal calculational errors.}
{\begin{tabular}{@{}ccccc@{}} \toprule
Program-Estimation method & Pioneer 10 (I) & Pioneer 10 (II) & Pioneer 10 (III) & Pioneer 11 \\ \colrule
{\footnotesize SIGMA}, WLS, no solar corona model & $8.02 \pm 0.01$ & $8.65 \pm 0.01$ & $7.83\pm0.01$ & $8.46\pm0.04$\\
{\footnotesize SIGMA}, WLS, with solar corona model & $8.00 \pm 0.01$ & $8.66\pm0.01$ & $7.84\pm0.01$ & $8.44\pm0.04$ \\
{\footnotesize SIGMA}, BSF, 1-day batch, with solar corona model & $7.82\pm0.29$ & $8.16\pm0.40$ & $7.59\pm0.22$ & $8.49\pm0.33$ \\
{\footnotesize CHASMP}, WLS, no solar corona model & $8.25\pm0.02$ & $8.86\pm0.02$ & $7.85\pm0.01$ & $8.71\pm0.03$ \\
{\footnotesize CHASMP}, WLS, with solar corona model & $8.22\pm0.02$ & $8.89\pm0.02$ & $7.92\pm0.01$ & $8.69\pm0.03$ \\
{\footnotesize CHASMP}, WLS, with corona, weighting, and F10.7 & $8.25\pm0.03$ & $8.90\pm0.03$ & $7.91\pm0.01$ & $8.91\pm0.04$ \\ \botrule
\end{tabular} \label{tab:1}}
\end{table}

We use a value near that quoted in table~I of Ref.~\refcite{Anderson} for the anomalous `acceleration' from interval III of the JPL analysis (12 July 1992 to 22 July 1998) of Pioneer 10 data to determine an approximate value for $\sqrt{\Lambda}$ in this theory since the solar radiation pressure on the spacecraft is most diminished in this interval (i.e. Pioneer 10 is farthest from the Sun in interval III; see fig.~6 of Ref.~\refcite{Anderson} for solar radiation pressure levels fitted during Earth-Jupiter cruise phase and Table~\ref{tab:1} in this work for a replication of table~I of Ref.~\refcite{Anderson}). We take $\sqrt{\Lambda} = 2.67 \times 10^{-18}$ s$^{-1}$ for the Hubble constant in our current analysis corresponding to an interval III anomalous `acceleration' for Pioneer 10 of approximately $a_p|_{\mathrm{P10}} = 7.84 \times 10^{-13}$ km s$^{-2}$ (eqn. (23) of Ref.~\refcite{Anderson}). We wish to emphasize that this article is concerned with the predicted behavior of the anomalous residuals after Saturn encounter for Pioneer 11 such that our focus is not primarily on the exact value of the anomalous `acceleration' seen by both spacecrafts. Meaning, we have used input data generated from {\footnotesize HORIZONS} to determine an approximate direction of propagation of these electromagnetic signals (i.e. to determine our $ |\hat{\mathbf k} \mathbf{\cdot} \hat{\mathbf r}|$ term) solely for the purposes of a rough calculation of residuals predicted by the theory of inertial centers. Nevertheless, we keep in mind that the Pioneer anomaly provides an avenue for determining the value of the Hubble constant within our own galaxy according to this theory. Fitting to the extended set of DSN frequency data in future work will help us to appropriately revise this value for $\sqrt{\Lambda}$.

\begin{figure}[]
\centerline{\psfig{file=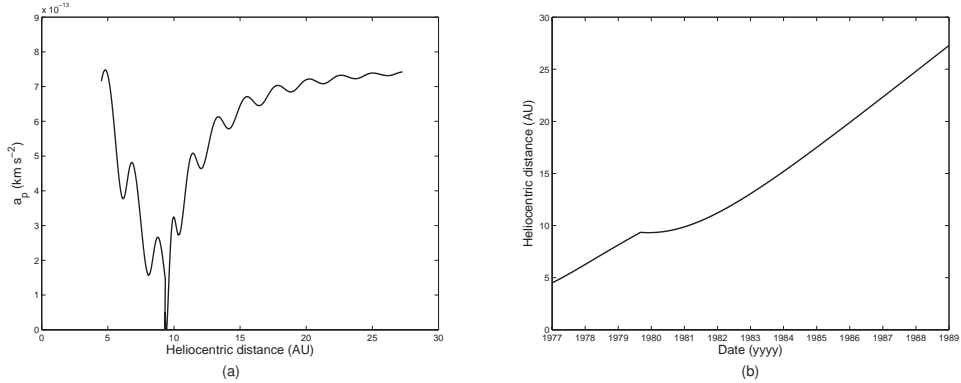,width=13.6cm}}
\vspace*{8pt}
\caption{Side-by-side plots of the anomalous `acceleration' predicted by the theory of inertial centers for Pioneer 11 assuming $\sqrt{\Lambda} = 2.67 \times 10^{-18}$ s$^{-1}$ (a) and the trajectory of the spacecraft in terms of heliocentric distance (b) from 1 January 1977 to 1 January 1989. Note that Saturn encounter occurred on 1 September 1979,\cite{NietoAnderson} subsequently resulting in a jump in the predicted anomalous `acceleration' as can be seen in (a) near a heliocentric distance of 10 AU. Input data used to generate these plots was taken from the JPL {\footnotesize HORIZONS} system. \label{fig:P11}}
\end{figure}

\begin{figure}[]
\centerline{\psfig{file=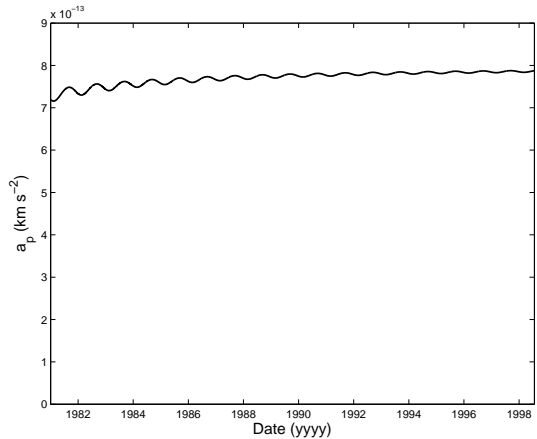,width=8.3cm}}
\caption{Anomalous `acceleration' predicted by the theory of inertial centers for Pioneer 10 from 1 January 1981 to 22 July 1998. \label{fig:P10}}
\end{figure}

Turning our attention to Fig.~\ref{fig:P11}(a) and Fig.~\ref{fig:P11}(b) where we have plotted the anomalous `acceleration' $a_p$ for Pioneer 11 according to the theory of inertial centers versus heliocentric distance, we realize by comparing with fig.~7 of Ref.~\refcite{Anderson} that the initial results from the predictions of this theory are promising. One finds near a heliocentric distance of 10 AU after encounter with Saturn by Pioneer 11 a jump in anomalous `acceleration' ultimately plateauing into a near-constant Hubble-like value for the clock acceleration as was observed and reported. Furthermore, we also find oscillations in the anomalous `acceleration' for the late-time behavior during plateau. We see from Fig.~\ref{fig:P10}, where we've plotted the predicted $a_p$ for Pioneer 10 versus date, that the theory of inertial centers prediction of a near-constant value for $a_p$ during the entire data interval plotted, where Pioneer 10 traverses a heliocentric distance range of approximately 20 to 70 AU, is consistent with the conclusions of section~V.A in Ref.~\refcite{Anderson} where the observed anomalous `acceleration' for Pioneer 10 was found to be within a range of $\pm 2 \times 10^{-13}$ km s$^{-2}$ of $(8.09 \pm 0.20) \times 10^{-13}$ km s$^{-2}$ for a spacecraft heliocentric distance between 40 and 60 AU. We also find oscillatory behavior in $a_p$ for Pioneer 10 as well with the oscillations suppressed over time.

It seems at first glance that this theory could provide a potential avenue to modeling the late-time annual oscillatory behavior discussed in section~IX.C of Ref.~\refcite{Anderson} as the near-annual period and decaying nature of the oscillations predicted for Pioneer 10 and Pioneer 11 in Fig.~\ref{fig:P11}(a) and Fig.~\ref{fig:P10}, respectively, are similar to that reported for Pioneer 10. Physically, one would provide the following reasoning behind these oscillations. As the trajectories of these spacecrafts were relatively radial in the galactic system during their cruise phases out of the Solar System, we would interpret the late-time oscillatory behavior of the anomalous `acceleration' seen by each as being due to Earth's revolution about the Sun each year, since this revolution produces an oscillatory change in the direction of propagation of our emitted photons relative to the galactic center (i.e. the factor $|\hat{\mathbf k} \mathbf{\cdot} \hat{\mathbf r}|$ oscillates throughout the year due to Earth's changing position in the Solar System). This, in turn, produces an oscillatory change in the amount the DSN's definition of time overlaps with physical clock-rates according to the theory of inertial centers. Furthermore, this predicted oscillation is suppressed over time since $\hat{\mathbf k}$ essentially becomes constant as our spacecrafts reach farther into interstellar space. Nevertheless, upon a slightly more detailed investigation of this behavior, we find that the sinusoidal predictions of this model for Pioneer 10 in interval III have amplitude an order of magnitude smaller than the quoted value of $(0.215 \pm 0.022) \times 10^{-13}$ km s$^{-2}$ from Ref.~\refcite{Anderson}. Using {\footnotesize HORIZONS} input data for Pioneer 10 from interval III, one finds with a simple nonlinear fit to an annual sinusoid in {\footnotesize MATLAB} an amplitude prediction of approximately $0.247 \times 10^{-14}$ km s$^{-2}$. While this most certainly does not rule out the possibility that the predictions of this theory can provide a viable alternative to the thermal explanation with the ability to model the onset as well as plateau for the anomalous `acceleration' of Pioneer 11, it does show that we may still need to attribute the observed oscillatory behavior to computer mismodeling as was originally suggested in Ref.~\refcite{Anderson}. The predicted oscillatory behavior shown in Fig.~\ref{fig:P10} during interval III for which the oscillatory analysis in Ref.~\refcite{Anderson} was conducted with {\footnotesize CHASMP} residuals would then likely be an order of magnitude too small to discern given the experimental precision for Pioneer 10 (see eqn.~(52) of Ref.~\refcite{Anderson}).

A separate source of concern in this theory's ability to provide a viable alternative to the thermal explanation for the Pioneer anomaly comes with its $a_p$ predictions for Pioneer 11 located at a heliocentric distance of less than 9 AU. Comparing with fig.~7 of Ref.~\refcite{Anderson}, one could initially find discrepancy between the observation of almost no anomalous `acceleration' at the first data point near 6 AU versus at least 50 percent of the total anomalous `acceleration' predicted by the theory of inertial centers in Fig.~\ref{fig:P11}(a). However, we must keep in mind the substantial effect that solar radiation pressure has on DSN frequency data at 6 AU relative to that of even a spacecraft heliocentric distance of 10 AU in our ability to discern a value for $a_p$. Referring to fig.~6 of Ref.~\refcite{Anderson}, we realize that, at a distance of 6 AU, the relatively large amount of solar radiation pressure could effectively drown out any anomalous effect at the level of $10^{-12}$ km s$^{-2}$ given the experimental precision for Pioneer 11. Consequently, the inconsistency between the first data point in fig.~6 of Ref.~\refcite{Anderson} and the theory of inertial centers predicted value in Fig.~\ref{fig:P11}(a) at 6 AU appears to be less of a concern than at first glance.

\begin{figure}[]
\centerline{\psfig{file=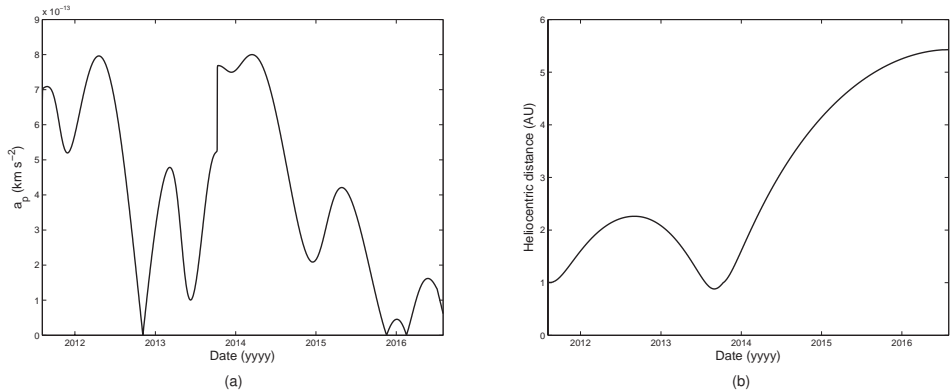,width=13.6cm}}
\caption{Side-by-side plots of the anomalous `acceleration' predicted by the theory of inertial centers for Juno (a) and the trajectory of the spacecraft in terms of heliocentric distance (b) from 6 August 2011 to 1 August 2016. Note that Earth flyby occurred on 9 October 2013.\cite{Thompson} \label{fig:Juno}}
\end{figure}

Focusing our attention on other possible detections of signatures of this theory with data from ongoing experiments, we turn to Juno and New Horizons. From Fig.~\ref{fig:Juno}(a), it is clear that, after Earth flyby on 9 October 2013, this model predicts an anomalous jump in `acceleration' for calculated DSN frequency residuals reminiscent of what one might expect from an effect resembling the flyby anomaly.\cite{Gravassist} However, before pursuing this idea further, we refer to the amplitude of these changes in `acceleration.' Given the data intervals used about Earth flyby are on the order of days instead of years along with the fact that light-times back and forth to the spacecraft are significantly smaller for Juno about Earth flyby than for Pioneer 10 and Pioneer 11 during the latter portion of their cruise phases out of the Solar System, we find that a change in `acceleration' on the order of $10^{-13}$ km s$^{-2}$ near this gravity assist date could not be detected with the fitting procedures employed even with a significant improvement in precision from Juno relative to that of Pioneer 10 and Pioneer 11.\cite{GravInversionJuno} This reasoning also applies for previous Earth flybys with other spacecrafts (see table~I in Ref.~\refcite{Gravassist}) where similar small changes are predicted by this model. It appears at first glance that the theory of inertial centers would be unable to explain the flyby anomaly unless some effect in our fitting procedures amplifies these small predicted changes in `acceleration.' However, a null signal prediction from this theory for Juno Earth flyby would be consistent with the recent analysis of Ref.~\refcite{Thompson} (see Ref.~\refcite{IorioJunoflyby} for a separate analysis of the flyby anomaly in the context of Juno Earth flyby).

\begin{figure}[]
\centerline{\psfig{file=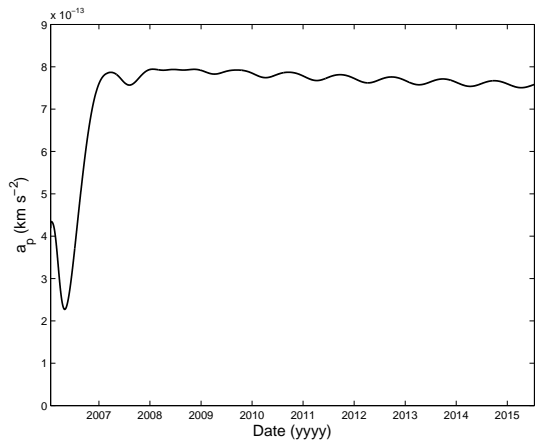,width=8.3cm}}
\caption{Anomalous `acceleration' predicted by the theory of inertial centers for New Horizons from 20 January 2006 to 1 September 2015. \label{fig:NewHorizons}}
\end{figure}

We look to cruise phase after Earth flyby for Juno, where the solar radiation pressure begins to decrease in approach to Jupiter. The relevant portion of Fig.~\ref{fig:Juno}a for us to consider for the predicted anomalous `acceleration' would be from early 2014 to early 2016. One finds from this figure a potential signature to decipher out of the solar radiation pressure between about 2 and 5 AU, referring to Fig.~\ref{fig:Juno}(b) for Juno heliocentric distance during this time interval. Our ability to discern a possible signal of the predictions presented in this article may be suppressed for Juno given the magnitude of solar radiation pressure inside of 6 AU. Nevertheless, DSN frequency data from cruise phase of New Horizons could provide an additional opportunity to test this theory because of the large decrease in solar radiation pressure outside of 10 AU relative to the magnitude of the `acceleration' that we are searching for as well as the nature of the signal predicted by this model, shown in Fig.~\ref{fig:NewHorizons}. From a spacecraft heliocentric distance of 10 AU in June 2008 to about 33 AU at anticipated Pluto encounter in July 2015, one finds an almost constant anomalous `acceleration' prediction nearly of the same magnitude as that observed with Pioneer 10.

\begin{table}[]
\tbl{Replication of table~II from the original Pioneer anomaly analysis.\cite{Anderson} Error budget for Pioneer 10 and Pioneer 11. A summary of biases and uncertainties.}
{\begin{tabular}{@{}cccc@{}} \toprule
Item & Description of error budget constituents & Bias ($\times 10^{-13}$ km s$^{-2}$) & Uncertainty ($\times 10^{-13}$ km s$^{-2}$) \\ \colrule
1 & Systematics generated external to the spacecraft: & & \\
 & (a) Solar radiation pressure and mass & $+0.03$ & $\pm 0.01$ \\
 & (b) Solar wind & & $\pm < 10^{-5}$ \\
 & (c) Solar corona & & $\pm 0.02$ \\
 & (d) Electro-magnetic Lorentz forces & & $\pm < 10^{-4}$ \\
 & (e) Influence of the Kuiper belt's gravity & & $\pm0.03$\\
 & (f) Influence of the Earth orientation & & $\pm0.001$ \\
 & (g) Mechanical and phase stability of DSN antennae & & $\pm < 0.001$ \\
 & (h) Phase stability and clocks & & $\pm < 0.001$ \\
 & (i) DSN station location & & $\pm < 10^{-5}$ \\
 & (j) Troposphere and ionosphere & & $\pm < 0.001$\\
2 & On-board generated systematics: & & \\
 & (a) Radio beam reaction force & $+1.10$ & $\pm0.11$ \\
 & (b) RTG heat reflected off the craft & $-0.55$ & $\pm0.55$ \\
 & (c) Differential emissivity of the RTGs & & $\pm0.85$ \\
 & (d) Non-isotropic radiative cooling of the spacecraft & & $\pm0.48$ \\
 & (e) Expelled Helium produced within the RTGs & +0.15 & $\pm0.16$ \\
 & (f) Gas leakage &  & $\pm0.56$ \\
 & (g) Variation between spacecraft determinations & +0.17 & $\pm0.17$ \\
3 & Computational systematics: & & \\
 & (a) Numerical stability of least-squares estimation & & $\pm 0.02$\\
 & (b) Accuracy of consistency and model tests & & $\pm0.13$\\
 & (c) Mismodeling of maneuvers & & $\pm0.01$\\
 & (d) Mismodeling of the solar corona & & $\pm0.02$\\
 & (e) Annual/diurnal terms & & $\pm0.32$\\
\colrule
 & Estimate of total bias or error & $+0.90$ & $\pm 1.33$ \\ \botrule
\end{tabular} \label{tab:2}}
\end{table}

\section{Conclusions}

To begin our conclusions, we offer remarks on the thermal explanation of the Pioneer anomaly. Refs.~\refcite{ThermalPioneer2011} and \refcite{ThermalPioneer2012} claim that the Pioneer anomaly can be explained by a thermal emission term in the line-of-sight acceleration of the spacecrafts that decays exponentially and is directed back toward the Solar System. This reasoning is consistent with the decay of the onboard $^{238}$Pu isotope by alpha emission with a half-life of 87.7 yr while taking into account the decrease with time in the efficiency of converting RTG power to spacecraft power. They conclude the Pioneer anomaly is 80 percent thermal based on telemetry data from Pioneer 10 and Pioneer 11 and further that there is enough uncertainty in the inference of thermal emission that a 100 percent contribution is not unlikely. Furthermore, their logic appears even more convincing after presenting fits to the extended set of Doppler tracking data. However, their conclusions are significantly different from those of Refs.~\refcite{TuryshevAnderson} and \refcite{Anderson} where thermal emission by the spacecrafts is taken to be 6 percent of the anomaly with an uncertainty of 6 percent. The analysis of Ref.~\refcite{Anderson} is based on blueprints of a SNAP-19 RTG power source with symmetrical and larger-than-usual cooling fins added specifically for the Pioneer spacecrafts. Each RTG was placed on one of two booms extending 3 m from the center of the spacecraft providing a total power delivered to the spacecraft bus of 155 W at launch. With a simplified spacecraft model, Ref.~\refcite{Anderson} suggests that a thermal emission of even 60 percent, the one-sigma value of Ref.~\refcite{ThermalPioneer2012}, is at the nine-sigma level. Additionally, the simple thermal model used in Ref.~\refcite{Anderson} was applied successfully in quite a different setting to the flyby of Cassini about the satellite Rhea,\cite{AndersonSchubert} so it does appear likely to be credible. One of the authors in Refs.~\refcite{Anderson} and \refcite{TuryshevAnderson} (ST) worked closely with another author (MMN) to develop Table~\ref{tab:2}, including the thermal model of 6\% with a realistic 6\% error bar. That author later lead the team that developed an alternative model based on Pioneer telemetry records. Although we doubt that the telemetry data can be stretched beyond a thermal contribution of 70\% $\pm$ 20\%, we nonetheless accept both published models as viable. A weighted mean of the two models yields 12.1\% $\pm$ 5.7\%, which we adopt as the best estimate of the thermal contribution.  The effect of the much larger estimate of the thermal contribution from the telemetry data is to positively bias the earlier estimate in Ref.~\refcite{Anderson} and Ref.~\refcite{TuryshevAnderson}, based on a detailed model of the Pioneer power subsystem, by one sigma.

A potential issue with the analysis of Ref.~\refcite{ThermalPioneer2012} arises if a best-fit inverse square with heliocentric distance term is subtracted from their predicted additional acceleration. {\it The result of subtracting such a term from their analysis is consistent with the claim by Ref.~\refcite{AndersonChameleon} that the anomalous acceleration appears to be nearly constant instead of decaying over time}. Such a term to subtract away is certainly present from solar radiation pressure, but Ref.~\refcite{ThermalPioneer2012} assume the existing solar pressure during cruise phase out of the Solar System for both Pioneer 10 and Pioneer 11 can be taken into account by using the model established during the cruise phase of the two spacecrafts {\it between Earth and Jupiter}. They do not introduce an extra $k/d_{\odot}^2$ term into their fitting model for spacecraft heliocentric distances farther than Jupiter, which is a reasonable assumption but nonetheless may not be correct. It is important to introduce $k$ as a free parameter and at least consider the possibility that it represents a difference in the model for solar radiation pressure between 1 and 5 AU with several parameters that could be biased by correlations and an appropriate model for solar radiation pressure with only one parameter in the region between 30 and 70 AU. In other words, we are not suggesting that Ref.~\refcite{ThermalPioneer2012} are incorrect in their analysis, but we claim instead that their results are not conclusive based on the telemetry data and the orbit determinations for the two spacecrafts. Thus, it would seem that the Pioneer anomaly remains a mystery until a new deep-space mission makes an unambiguous measurement of any extra acceleration or lack thereof, or until a new theory predicts the existence of the extra acceleration observed with Pioneer 10 and Pioneer 11 in a manner consistent with all of the observed features of the anomaly. Our purpose with this article is to indicate a possible theoretical direction which for now seems to be consistent with the observed anomaly.

We attempted to model the possible onset of the Pioneer anomaly after Pioneer 11 Saturn encounter using predictions from the theory of inertial centers to offer a potential alternative to the thermal explanation of Ref.~\refcite{ThermalPioneer2012}. We derived the form of the frequency residuals calculated by the DSN according to this theory and discovered a drift term with a clock acceleration similar in magnitude to that of Ref.~\refcite{Kopeikin} but having an additional weighting factor dependent upon the direction of transmission of each photon relative to the direction of our galactic center. However, an ambiguity arose with regard to the question of how to relate time as defined by the DSN to physical clock-rates in the theory of inertial centers. We adopted a particular convention to address the ambiguity that allowed both time coordinates to span a ``physical range" (i.e. $0<T_{\mathrm{DSN}}<\infty$ and $0<t<\infty$) and, with this convention, found that our DSN residual drift term is blueshifted in nature for both Pioneer 10 and Pioneer 11 as observed in Ref.~\refcite{Anderson}. After plotting with {\footnotesize HORIZONS} input data, it became clear that our additional weighting factor provided the necessary adjustment to produce a jump in anomalous `acceleration' after Pioneer 11 Saturn flyby with an eventual plateau causing a constant drift of DSN fractional frequency residuals with rate of change similar to the value of the Hubble constant.

Nevertheless, the additional `frequency shift' according to this cosmological model is not in fact a physical shift in frequency of each of these photons. Instead, due to the manner in which the DSN calculates changes in frequency, we interpreted this predicted shift as a manifestation of the difference in clock-rates between the DSN's definition of time (\ref{Minktime}) and time according to the theory of inertial centers. As a consistency check, we also found our results for Pioneer 10 coincided with the observations reported in Ref.~\refcite{Anderson} as a near-constant clock acceleration for the set of data analyzed. Thus, prior to proper fitting of this model to the extended set of DSN frequency data mentioned in our introduction, it appears that the predictions of the theory of inertial centers could potentially provide an alternative to the thermal explanation of the Pioneer anomaly with the ability to replicate an onset of said anomaly after Saturn encounter by Pioneer 11. Still, we are far from satisfied with our explanation for why our convention choice somehow resolves the mathematical ambiguity one encounters from the restrictions imposed by this theory. In particular, given that our restrictions on the DSN time coordinate in our latter scenario (\ref{latter}) are physically consistent with the expectations from the DSN's definition of range, why should our choice in adopted convention have any physical relevance at all? And why can we consider the latter scenario to be unphysical? A completely satisfactory resolution of this matter has yet to be determined.

Future work with regard to the issue of the possible onset of the Pioneer anomaly requires fitting of the additional clock acceleration term predicted by the theory of inertial centers to the residuals of the extended set of DSN frequency data for Pioneer 10 and Pioneer 11. Accurate determination of $\sqrt{\Lambda}$ should result from this analysis once adjustments to the expected DSN residuals from both spacecrafts are incorporated. These adjustments necessary for fitting include the revised thermal model for the spacecrafts mentioned earlier in our conclusions with a weighted contribution estimate to the anomaly of 12.1\% $\pm$ 5.7\% as well as any adjustments to models for the solar radiation pressure such as those suggested in Ref.~\refcite{AndersonChameleon}. After removing these contributions from the Pioneer anomaly signal in the DSN residuals, we should find a fitted value for the Hubble constant according to the theory of inertial centers from both Pioneer data sets and subsequently be able to directly quantify the goodness of fit of the predictions of the theory of inertial centers to actual observations. Additionally, we hope to fit possible releases of DSN frequency data from the cruise phases of Juno and New Horizons to the predictions displayed in Figs.~\ref{fig:Juno}a and \ref{fig:NewHorizons} to further test this model against observation.

As remarked upon at the end of \ref{gendopplershift}, a separate line of inquiry into the ability of this theory to resolve other observed Solar System anomalies seems to appear when incorporating the relative velocity of our observers in DSN-calculated fractional frequency residuals. We find another DSN Doppler term that one would likely attribute to a fictitious `acceleration'
\begin{equation}\label{iorioextra}
{\mathbf a}_{\mathrm{pert}} = 2 |\hat{\mathbf k}_1 \mathbf{\cdot} \hat{\mathbf r}|\sqrt{\Lambda} {\mathbf v}_R,
\end{equation}
where ${\mathbf v}_R$ is the line-of-sight velocity between our observers. This `acceleration' term is of a similar form as the empirical solution proposed in Ref.~\refcite{IorioAULunar} to resolve the anomalous secular increases of the astronomical unit and the eccentricity of the lunar orbit. These predictions from (\ref{iorioextra}) could become apparent with the analysis of Ref.~\refcite{Krasinsky2004} for the secular increase in the astronomical unit as their calculations employ significantly more Doppler measurements than the ephemerides, although apparently ``downweighted to avoid influence of their systematic errors" (see their section~3.1). While it is clear from our earlier calculations that the predictions of the theory of inertial centers would not affect ranging measurements to the Moon, there are subtleties that need to be investigated further with regard to reported increases in the eccentricity of the lunar orbit as determinations of lunar $GM_{\mathrm{Moon}}$ rely upon DSN Doppler data from orbiting spacecrafts.\cite{Williams_lunarcore, Konopliv1998, Konopliv2001} On the other hand, if (\ref{iorioextra}) were able to predict the anomalous increases in the AU and lunar orbit eccentricity, why then would our larger clock-drift term not be apparent in these analyses?

\section*{Acknowledgments}

This work was supported by NASA contract NNM06AA75C for the Juno mission.

\appendix

\section{Geodesic solutions}\label{GeodesicSolutions}

We look for geodesic solutions in a particular inertial frame of reference as defined by the theory of inertial centers.\cite{Feldman} Our geodesic equation is given by
\begin{equation}
0 = U^a \nabla_a U^b,
\end{equation}
where $U^{\mu} = dx^{\mu}/d\sigma$ are the components of the affinely parametrized tangent vector to our geodesic and $x^{\mu}(\sigma)$ is our coordinate parametrization of the geodesic. To find solutions, we start by expressing this equation in terms of Minkowski coordinates $(T, {\mathbf X})$, where we take the spatial coordinate origin of this system to be located at the position of our inertial center (i.e. $r=0 \rightarrow \langle X,Y,Z \rangle =\langle 0,0,0\rangle$). Our metric in terms of this Minkowski coordinate parametrization of our inertial reference frame in this theory takes the form
\begin{equation}
ds^2 = -c_0^2 dT^2 + \sum_{i} dX_i^2,
\end{equation}
where $i =1,2,3$, $X^1 = X, X^2 = Y, X^3 = Z$, and $ds^2 = -d\chi^2 = -\Lambda r^2 d\tau^2$ for massive geodesics. $r(\tau)$ is the physical distance to our inertial center. As our metric components in Minkowski coordinates are all constants, our affine connection tensor for Minkowski coordinates is zero: $\Gamma^{c}_{a b}|_{\rm Minkowski} =0$. Our equations of motion in Minkowski coordinates are then found to be
\begin{eqnarray}
\frac{d^2 T}{d\sigma^2} = 0
\\ \frac{d^2 X^i}{d \sigma^2} = 0,
\end{eqnarray}
where $\sigma \rightarrow \chi = \sqrt{\Lambda} \int r(\tau) d\tau$ in the massive case. Examining only spatial components, we see that
\begin{eqnarray}
0 = \frac{d^2X^i}{d\sigma^2} = \frac{d}{d\sigma}\bigg(\frac{dT}{d\sigma}\frac{dX^i}{dT}\bigg) \nonumber \\
= \frac{d^2 T}{d \sigma^2}\frac{dX^i}{dT} + \bigg(\frac{dT}{d\sigma}\bigg)^2 \frac{d^2X^i}{dT^2},
\end{eqnarray}
which leads us to
\begin{equation} \label{diffMinkgeo}
\frac{d^2 X^i}{dT^2} = 0,
\end{equation}
where we have used the equation of motion for our Minkowski time coordinate $T$.

The solutions to (\ref{diffMinkgeo}) are of the form
\begin{equation} \label{Minkgeo}
X^i (T) = V^i T + X^i_0,
\end{equation}
where $X^i_0$ is our spatial Minkowski coordinate location when $T=0$ and $V^i = dX^i/dT$ are independent of $T$ restricted by the normalization condition $\sum_i (V^i)^2 \leq c_0^2$. Nevertheless, we know from the form of our affine parameter $\chi$ that we wish to express this coordinate parametrization of our geodesics in terms of the more physically applicable radial Rindler coordinates where coordinate time $t$ in the radial Rindler chart (see (\ref{metric})) progresses at the same rate as a physical clock of an observer stationary relative to our inertial center according to this theory (i.e. Minkowski coordinate time $T$ does {\it not} correspond to a physical clock-rate). The relevant coordinate transformations from our Minkowski coordinate parameterization are
\begin{eqnarray}
T = \frac{r}{c_0}\sinh(\sqrt{\Lambda}t) \\
X = r \cosh(\sqrt{\Lambda}t) \sin{\theta} \cos{\phi} \\
Y = r \cosh(\sqrt{\Lambda}t) \sin{\theta} \sin{\phi} \\
Z = r \cosh(\sqrt{\Lambda}t) \cos{\theta}.
\end{eqnarray}
Using these transformations in (\ref{Minkgeo}), we find
\begin{eqnarray}
r \cosh(\sqrt{\Lambda}t) \sin{\theta} \cos{\phi} = \bigg(\frac{V^X}{c_0}\bigg) r\sinh(\sqrt{\Lambda}t) + X_0 \\
r \cosh(\sqrt{\Lambda}t) \sin{\theta} \sin{\phi} = \bigg(\frac{V^Y}{c_0}\bigg) r\sinh(\sqrt{\Lambda}t) + Y_0 \\
r \cosh(\sqrt{\Lambda}t) \cos{\theta} = \bigg(\frac{V^Z}{c_0}\bigg) r\sinh(\sqrt{\Lambda}t) + Z_0.
\end{eqnarray}
Solving for $r$ by squaring each one of these equations and adding the results together
\begin{equation}
r^2 \cosh^2(\sqrt{\Lambda}t) = \sum_{i} \bigg(\frac{V^i}{c_0} r\sinh(\sqrt{\Lambda}t) + X^i_0 \bigg)^2,
\end{equation}
where we choose our time coordinates $T$ and $t$ to coincide for our initial conditions (i.e. $t=0 \iff T=0$ and $[dt = dT]|_{t=T=0}$). This requirement forces our Minkowski constant $c_0$ to be equal to $\sqrt{\Lambda}r_0$ where $r_0$ is the initial distance of our object to the inertial center and $\sqrt{\Lambda}$ is fixed by the value of the Hubble constant.

We employ the relations
\begin{eqnarray}
\sum_i (X^i_0)^2 = \sum_i (X^i(T))^2|_{T=0} = \sum_i (X^i(t))^2|_{t=0} = r_0^2
\\ RV^{R} = \sum_i X^i \frac{dX^i}{dT} = \sum_i X^i V^i,
\end{eqnarray}
where $R^2 = \sum_i (X^i)^2$ and $V^R(T) = dR/dT$. Yet, since $dX^i/dT$ is a constant for all $T$, we have
\begin{equation}
\sum_i X_0^i V^i = RV^R|_{T=0} = Rv^R |_{t=0} = r_0 v_0^r,
\end{equation}
where $v_0^r = dr/dt|_{t=0}$ is the initial radial velocity of the object in the inertial system (i.e. $v^i = dx^i/dt$ are physical velocities). Plugging in above, we find after solving for $r(t)$
\begin{eqnarray}
r(t) = \frac{r_0}{1 + \sinh^2(\sqrt{\Lambda}t) (1 - \sum_{i} (\frac{V^i}{c_0})^2 )} \bigg\{\frac{v_0^r}{c_0}\sinh(\sqrt{\Lambda}t) \nonumber
\\ \pm \sqrt{\bigg(\frac{v_0^r}{c_0}\bigg)^2 \sinh^2(\sqrt{\Lambda}t) + \bigg[1 + \sinh^2(\sqrt{\Lambda}t) \bigg(1 - \sum_{i} \bigg(\frac{V^i}{c_0}\bigg)^2 \bigg)\bigg] } \bigg\}.
\end{eqnarray}
When $t=0$, we want our expression to reduce to $r_0$ and not $-r_0$. Therefore, we take only the positive root for our solution.

Additionally, we see that $\sum_{i} (V^i)^2$ can be expressed in terms of the induced Riemannian metric components $\tilde{h}_{ij}$ for a space-like hypersurface in our original Minkowski coordinates in the following manner:
\begin{equation}
\sum_{i} (V^i)^2 = \sum_{i,j} \tilde{h}_{ij} V^i V^j = v_0^2.
\end{equation}
However, this is simply the squared magnitude of the $V^i$'s in Minkowski coordinates along this hypersurface. Furthermore, since our time coordinates coincide for our initial conditions, $v_0$ can be interpreted physically as the initial speed of our inertially traveling object measured in terms of the clock of a stationary observer located at $r=r_0$. In other words, since $V^i = dX^i/dT = dX^i/dT|_{T=0}$,
\begin{eqnarray}
v_0^2 = \sum_{i,j} \tilde{h}_{ij} \frac{dX^i}{dT} \frac{dX^j}{dT} = \sum_{i,j} \tilde{h}_{ij} \frac{dX^i}{dT} \frac{dX^j}{dT} \bigg|_{T=0} \nonumber \\
= \sum_{i,j} \sum_{k,l} \tilde{h}_{ij} \frac{\partial X^i}{\partial x^k}\frac{d x^k}{dT} \frac{\partial X^j}{\partial x^l} \frac{dx^l}{dT}\bigg|_{T=0} \nonumber \\
= \sum_{k,l} \bigg(\sum_{i,j} \tilde{h}_{ij}\frac{\partial X^i}{\partial x^k} \frac{\partial X^j}{\partial x^l} \bigg)\frac{d x^k}{dt}\frac{d x^l}{dt}\bigg|_{t=0} \nonumber \\
= \sum_{k,l} h_{kl} \frac{dx^k}{dt} \frac{dx^l}{dt}\bigg|_{t=0} = \sum_{k,l} h_{kl} v^k v^l |_{t=0} \nonumber
\\ = (v_0^r)^2 + r_0^2\bigg((v_0^{\theta})^2 + \sin^2{\theta_0}(v_0^{\phi})^2 \bigg),
\end{eqnarray}
where $h_{kl}$ are induced Riemannian metric components along the $t=T=0$ hypersurface for the radial Rindler chart and $x^k$ are spatial coordinates in the same chart. Plugging back into our expression for $r(t)$ where we remember to take only the positive root, we find for the radial position of our object
\begin{eqnarray} \label{radialmotion}
r(t) = \frac{r_0}{1 + \sinh^2(\sqrt{\Lambda}t) (1 - (\frac{v_0}{c_0})^2 )} \bigg\{\frac{v_0^r}{c_0}\sinh(\sqrt{\Lambda}t) \nonumber
\\ + \sqrt{1 + \sinh^2(\sqrt{\Lambda}t) \bigg[1 + \bigg(\frac{v_0^r}{c_0}\bigg)^2 - \bigg(\frac{v_0}{c_0}\bigg)^2\bigg] } \bigg\},
\end{eqnarray}
with $c_0 = \sqrt{\Lambda}r_0$.

As briefly alluded to earlier in this section, our normalization condition for the tangent vector to a geodesic is
\begin{equation}
- k = g_{ab}U^a U^b,
\end{equation}
where
\begin{equation}
k = \left\{\begin{array}{l l} 0 & \quad \textrm{null geodesics} \\ 1 & \quad \textrm{time-like geodesics} \\ \end{array}\right..
\end{equation}
We can express this normalization condition in our original Minkowski coordinates in the following manner:
\begin{eqnarray}
- k = -c_0^2 \bigg(\frac{dT}{d\sigma} \bigg)^2 + \sum_i \bigg(\frac{dX^i}{d\sigma} \bigg)^2 \nonumber
\\ = -\bigg(\frac{dT}{d\sigma}\bigg)^2 \bigg[c_0^2 - \sum_i (V^i )^2 \bigg] = -\bigg(\frac{dT}{d\sigma}\bigg)^2 \bigg[c_0^2 - v_0^2 \bigg],
\end{eqnarray}
which provides us with the constraints
\begin{eqnarray}
v_0 = c_0 \indent \textrm{massless objects} \label{masslesslightspeed} \\
v_0 < c_0 \indent \textrm{massive objects}.
\end{eqnarray}
Returning to our coordinate transformations, the angular positions of our object as a function of inertial time $t$ are found to be
\begin{eqnarray}
\phi(t) = \tan^{-1}\bigg[\frac{\frac{v^y_0}{c_0} r(t) \sinh(\sqrt{\Lambda}t) + Y_0}{\frac{v^x_0}{c_0} r(t) \sinh(\sqrt{\Lambda}t) + X_0} \bigg] \nonumber \\
= \phi_0 + \tan^{-1}\bigg[\frac{r\sinh(\sqrt{\Lambda}t) \frac{v_0^{\phi}}{c_0} r_0 \sin{\theta_0}}{r_0 \sin{\theta_0} + r\sinh(\sqrt{\Lambda}t)[\frac{v_0^r}{c_0} \sin{\theta_0} + \frac{v_0^{\theta}}{c_0} r_0 \cos{\theta_0}]} \bigg],
\end{eqnarray}
and
\begin{eqnarray}
\theta (t) = \cos^{-1}\bigg[\frac{v^z_0}{c_0}\tanh(\sqrt{\Lambda}t) + \frac{Z_0}{r(t)\cosh(\sqrt{\Lambda}t)} \bigg] \nonumber \\
= \cos^{-1}\bigg[\tanh(\sqrt{\Lambda}t) \bigg(\frac{v_0^r}{c_0}\cos{\theta_0} - \frac{v_0^{\theta}}{c_0}r_0 \sin{\theta_0} \bigg) + \frac{r_0 \cos{\theta_0}}{r \cosh(\sqrt{\Lambda}t)} \bigg],
\end{eqnarray}
where we've employed
\begin{eqnarray}
X(t)|_{t=0} = X_0 = r_0 \sin{\theta_0}\cos{\phi_0} \\
Y(t)|_{t=0} = Y_0 = r_0 \sin{\theta_0}\sin{\phi_0} \\
Z(t)|_{t=0} = Z_0 = r_0 \cos{\theta_0},
\end{eqnarray}
and
\begin{eqnarray}
V^X = v_0^x = v_0^r \sin{\theta_0} \cos{\phi_0} + v_0^{\theta} r_0 \cos{\theta_0}\cos{\phi_0} - v^{\phi}_0 r_0 \sin{\theta_0} \sin{\phi_0}\label{v1} \\
V^Y = v_0^y = v_0^r \sin{\theta_0} \sin{\phi_0} + v_0^{\theta} r_0 \cos{\theta_0}\sin{\phi_0} + v^{\phi}_0 r_0 \sin{\theta_0} \cos{\phi_0}\label{v2} \\
V^Z = v_0^z = v_0^r \cos{\theta_0} - v_0^{\theta} r_0 \sin{\theta_0}.\label{v3}
\end{eqnarray}

\section{DSN frequency shift with the Doppler effect}\label{gendopplershift}

To factor in the Doppler shift from the relative velocity of our observers, we return to the general expression for the observed wavelength of a photon according to the theory of inertial centers taken in Ref.~\refcite{Feldman} to be
\begin{equation}
-\frac{2\pi}{\lambda} = k_a U^a \bigg|_{P},
\end{equation}
where $U^a$ is the `four-velocity' of the observer affinely paramterized by $\chi$ (not simply $\tau$) and given by $U^{\mu} = dx^{\mu}/d\chi$ in component form, $k^a$ is the wave-vector of the photon, and $P$ refers to the point of emission/reception of the electromagnetic signal in question. We then have the following relations as the observer is assumed to follow a time-like geodesic while the photon propagates along a null geodesic:
\begin{eqnarray}
U^aU_a = -1
\\ k^a k_a = 0.
\end{eqnarray}
Expanding these relations in the radial Rindler chart, one finds
\begin{equation}
U^t = \bigg[\Lambda r^2 - (v^r)^2 - r^2 \cosh^2(\sqrt{\Lambda}t)\bigg((v^{\theta})^2 + \sin^2{\theta} (v^{\phi})^2 \bigg)\bigg]^{-1/2},
\end{equation}
\begin{equation}
\Lambda r^2 = (c^r)^2 + r^2 \cosh^2(\sqrt{\Lambda}t)\bigg((c^{\theta})^2 + \sin^2{\theta}(c^{\phi})^2\bigg),
\end{equation}
where we define $v^{i} \equiv dx^i/dt$ and $c^i \equiv k^i/k^t$ employing the Latin index $i$ in the component form of these vectors to refer to spatial components. Notice that $v^{i}$ is the local velocity of our observer {\it relative to the inertial center about which he/she moves} (the center of the Milky Way for our purposes). Yet, along a constant $t$ spatial hypersurface with induced Riemannian metric components $h_{ij}$, one finds
\begin{eqnarray}
c^2 = \sum_{i} c^i c_i = \sum_{i,j} h_{ij}c^i c^j  \nonumber
\\ = (c^r)^2 + r^2 \cosh^2(\sqrt{\Lambda}t) \bigg( (c^{\theta})^2 + \sin^2{\theta} (c^{\phi})^2 \bigg),
\end{eqnarray}
and thus
\begin{equation}
\sum_{i}c^i c_i = \Lambda r^2.
\end{equation}
Note that $c$ here is {\it not} the DSN-adopted value for the speed of light in empty vacuum. Instead $c$ in this section refers to the speed of the photon at a particular space-time point in the inertial frame according to this theory. For our own ease, we also denote the square of the local velocity of our observer relative to the center of the inertial system by
\begin{equation}
v^2 = \sum_i v^i v_i = (v^r)^2 + r^2 \cosh^2(\sqrt{\Lambda}t) \bigg( (v^{\theta})^2 + \sin^2{\theta} (v^{\phi})^2 \bigg).
\end{equation}
Then our expression for $U^t$ can be written in a simpler form,
\begin{equation}\label{timefour}
U^t = \frac{1}{\sqrt{\Lambda}r}\cdot \frac{1}{\sqrt{1 - v^2/c^2}},
\end{equation}
and the wavelength measured by our observer is found to be
\begin{equation}
-\frac{2\pi}{\lambda} = - \frac{\sqrt{\Lambda}r}{\sqrt{1- v^2/c^2}} k^t \bigg[1 - \frac{\sum_{i} v^i c_i}{c^2} \bigg],
\end{equation}
where $\sum_{i}v^i c_i = \sum_{i,j}h_{ij}v^i c^j$ can be thought of as the `dot-product' of the observer and photon velocities relative to the center of the inertial system along the constant $t$ spatial hypersurface. To find an expression for $k^t$, we make use of the Killing vector field $\xi^a = (1/\sqrt{\Lambda}r)\cosh(\sqrt{\Lambda}t)(\partial/\partial t)^a - \sinh(\sqrt{\Lambda}t) (\partial/\partial r)^a$ in the conservation equation $-E = \xi^a k_a$, where $E$ is a constant.\cite{Feldman} One finds
\begin{equation}
k^t = \frac{E}{\sqrt{\Lambda}r [\cosh(\sqrt{\Lambda}t) + (c^r/\sqrt{\Lambda}r)\sinh(\sqrt{\Lambda}t)]},
\end{equation}
and our relation for the observed wavelength reduces to
\begin{eqnarray} \label{onewaywavelength}
\frac{2\pi}{\lambda} = \frac{E}{\cosh(\sqrt{\Lambda}t) + (c^r/\sqrt{\Lambda}r)\sinh(\sqrt{\Lambda}t)} \frac{1}{\sqrt{1-v^2/c^2}}\bigg[1- \frac{\sum_i v^i c_i}{c^2} \bigg].
\end{eqnarray}

We recall $U^t = dt/d\chi = (1/\sqrt{\Lambda}r)\cdot dt/d\tau$, where $\tau$ is the proper time according to the theory of inertial centers. Comparing this expression for $U^t$ with (\ref{timefour}) and plugging into (\ref{onewaywavelength}), we have
\begin{eqnarray}
\frac{2\pi}{\lambda} = \frac{E}{\cosh(\sqrt{\Lambda}t) + (c^r/\sqrt{\Lambda}r)\sinh(\sqrt{\Lambda}t)} \cdot \frac{dt}{d\tau} \cdot \bigg[1 - \frac{\sum_i v^i c_i}{c^2} \bigg].
\end{eqnarray}
Taking the ratio of observable wavelengths, one finds for the uplink shift
\begin{eqnarray}
\frac{\lambda_1}{\lambda_2}\bigg|_{\rm uplink} = \frac{\cosh(\sqrt{\Lambda}t_1) + (c_1^r/\sqrt{\Lambda}r_1)\sinh(\sqrt{\Lambda}t_1)}{\cosh(\sqrt{\Lambda}t_2) + (c_2^r/\sqrt{\Lambda}r_2)\sinh(\sqrt{\Lambda}t_2)} \cdot \frac{(dt/d\tau)_2}{(dt/d\tau)_1} \nonumber
\\ \cdot \bigg[1 - \frac{\sum_i v^i c_i}{c^2} \bigg]\bigg|_2 \bigg[1 - \frac{\sum_i v^i c_i}{c^2} \bigg]^{-1} \bigg|_1,
\end{eqnarray}
and for the downlink
\begin{eqnarray}
\frac{\lambda_2}{\lambda_3}\bigg|_{\rm downlink} = \frac{\cosh(\sqrt{\Lambda}t_2) - (c_2^r/\sqrt{\Lambda}r_2)\sinh(\sqrt{\Lambda}t_2)}{\cosh(\sqrt{\Lambda}t_3) + (c_3^r/\sqrt{\Lambda}r_3)\sinh(\sqrt{\Lambda}t_3)}\cdot \frac{(dt/d\tau)_3}{(dt/d\tau)_2} \nonumber
\\ \cdot \bigg[1- \frac{\sum_i v^i c_i}{c^2} \bigg]\bigg|_{3} \bigg[1 + \frac{\sum_i v^ic_i}{c^2} \bigg]^{-1}\bigg|_{2},
\end{eqnarray}
where we have implicitly taken $c^i_2|_{\rm downlink} = - c^i_2|_{\rm uplink}$ and $v^i_2|_{\rm uplink} = v^i_2|_{\rm downlink}$. Furthermore, all quantities marked with a `2' in the rest of our analysis refer to the uplink portion of the two-way traversal, and we assume that $\hat{\mathbf k}_1 = \hat{\mathbf k}_2|_{\rm uplink} = - \hat{\mathbf k}_2|_{\rm downlink} = - \hat{\mathbf k}_3$, $v^i|_{1} = v^i|_{3}$, and $r_1 = r_3$ which implies $c_1 = c_3$. Our expression for the two-way shift then reduces to
\begin{eqnarray}
\frac{\lambda_1}{\lambda_3} = \frac{\cosh(\sqrt{\Lambda}t_1) + (c_1^r/\sqrt{\Lambda}r_1)\sinh(\sqrt{\Lambda}t_1)}{\cosh(\sqrt{\Lambda}t_3) - (c_1^r/\sqrt{\Lambda}r_1)\sinh(\sqrt{\Lambda}t_3)} \cdot \frac{\cosh(\sqrt{\Lambda}t_2) - (c_2^r/\sqrt{\Lambda}r_2)\sinh(\sqrt{\Lambda}t_2)}{\cosh(\sqrt{\Lambda}t_2) + (c_2^r/\sqrt{\Lambda}r_2)\sinh(\sqrt{\Lambda}t_2)} \nonumber \\
 \cdot \frac{(dt/d\tau)_3}{(dt/d\tau)_1} \cdot \frac{1+(\sum_i v^i c_i/c^2)_1}{1-(\sum_i v^i c_i/c^2)_1} \cdot \frac{1- (\sum_i v^i c_i/c^2)_2}{1 + (\sum_i v^i c_i/c^2)_2}.
\end{eqnarray}
Simplifying the first line of the above expression with Taylor series expansions up to first-order in $\sqrt{\Lambda}t$, we have
\begin{eqnarray}
\frac{\lambda_1}{\lambda_3} =  \bigg[1 + (\hat{\mathbf k}_1 \cdot \hat{\mathbf r})\sqrt{\Lambda}[(t_3 - t_2) - (t_2 - t_1)] + \mathcal{O}(\Lambda t^2) \bigg] \cdot \frac{(dt/d\tau)_3}{(dt/d\tau)_1}\nonumber
\\ \cdot \frac{1+(\sum_i v^i c_i/c^2)_1}{1-(\sum_i v^i c_i/c^2)_1} \cdot \frac{1- (\sum_i v^i c_i/c^2)_2}{1 + (\sum_i v^i c_i/c^2)_2},
\end{eqnarray}
where we have substituted for $c^r_1 = c_1 (\hat{\mathbf k}_1 \mathbf{\cdot} \hat{\mathbf r}) =  \sqrt{\Lambda}r_1 (\hat{\mathbf k}_1 \mathbf{\cdot} \hat{\mathbf r})$ and $c^r_2 = c_2 (\hat{\mathbf k}_2|_{\rm uplink} \mathbf{\cdot} \hat{\mathbf r}) =  \sqrt{\Lambda}r_2 (\hat{\mathbf k}_2|_{\rm uplink} \mathbf{\cdot} \hat{\mathbf r})= \sqrt{\Lambda}r_2 (\hat{\mathbf k}_1 \mathbf{\cdot} \hat{\mathbf r})$, taking the galactic radial unit vector to be approximately the same for our observers in the Solar System (i.e. $\hat{\mathbf r}_1 = \hat{\mathbf r}_2 = \hat{\mathbf r}$). Yet from our earlier analysis in the discussion of this article, we know that the elapsed coordinate time for the uplink, $t_{\rm uplink}$, will be approximately equal to the elapsed coordinate time for the downlink, $t_{\rm downlink}$. In other words, $t_3 - t_2 = t_2 - t_1$ resulting in
\begin{eqnarray}\label{genshift}
\frac{\lambda_1}{\lambda_3} = \frac{(dt/d\tau)_3}{(dt/d\tau)_1} \cdot \frac{1+(\sum_i v^i c_i/c^2)_1}{1-(\sum_i v^i c_i/c^2)_1} \cdot \frac{1- (\sum_i v^i c_i/c^2)_2}{1 + (\sum_i v^i c_i/c^2)_2} + \mathcal{O}(\Lambda t^2).
\end{eqnarray}

However, for relative frequency shifts calculated by the DSN, one must realize that the DSN defines physical clock-rates from relation (\ref{Minktime}). Thus, while our actual fractional change in frequency once formulating in terms of coordinate time corresponding with physical stationary clock-rates according to the theory of inertial centers will not produce a drift term as shown later in this section, {\it what the DSN calculates in terms of the clock-rates it assumes to be physical will}. In other words, the range-rate calculated by the DSN uses the DSN's definition of time for its time coordinate. However, the DSN's time coordinate drifts away from actual physical time according to the theory of inertial centers, and thus with this DSN calculation we no longer have an accurate assessment of the frequency observable itself (i.e. $\nu_3/\nu_1 \neq [\nu_3/\nu_1]|_{\rm DSN}$). To understand what the DSN calculates, it appears we must substitute $t \rightarrow T_{\rm DSN}$ in our previous expression as $T = T_{\rm DSN}$ defined by (\ref{Minktime}) is the coordinate time that the DSN uses in its range-rate calculation. Still, in our previous expression, $\tau$ remains the physical proper time according to the theory of inertial centers. We also keep in mind that $v^i = dx^i/dt$ is the local velocity of the observer in the inertial frame. This local velocity happens to coincide with the velocity expected from the DSN when examining for our slow-moving massive objects (i.e. $dr/d\tau \approx 0$ in this limit in (\ref{geo})). Then,
\begin{eqnarray}
\bigg[\frac{\lambda_1}{\lambda_3}\bigg]_{\rm DSN} = \frac{(dT/d\tau)_3}{(dT/d\tau)_1} \cdot \frac{1+(\sum_i v^i c_i/c^2)_1}{1-(\sum_i v^i c_i/c^2)_1} \cdot \frac{1- (\sum_i v^i c_i/c^2)_2}{1 + (\sum_i v^i c_i/c^2)_2} + \mathcal{O}(\Lambda t^2).
\end{eqnarray}
If we keep terms up to first-order in $(v/c)$, we see from our earlier expression for $U^t$ in (\ref{timefour}) along with the form of our affine parameter in (\ref{eq:affine}) that
\begin{equation}
\frac{dt}{d\tau} = 1 + \mathcal{O}[(v/c)^2].
\end{equation}
Thus,
\begin{eqnarray}
\bigg[\frac{\lambda_1}{\lambda_3}\bigg]_{\rm DSN} = \frac{(dT/dt)_3}{(dT/dt)_1} \cdot \frac{1+(\sum_i v^i c_i/c^2)_1}{1-(\sum_i v^i c_i/c^2)_1} \cdot \frac{1- (\sum_i v^i c_i/c^2)_2}{1 + (\sum_i v^i c_i/c^2)_2} + \mathcal{O}[\Lambda t^2, (v/c)^2].
\end{eqnarray}

Finally, we use our expression for the difference in clock-rates between time according to the theory of inertial centers and time assumed by the DSN. From (\ref{drift1}), we have
\begin{equation}
\bigg(\frac{dT}{dt}\bigg)_i = \frac{\sqrt{\Lambda}r_s}{c_{\rm DSN}}\bigg[1 + \sqrt{\Lambda}|\hat{\mathbf k}_i \mathbf{\cdot} \hat{\mathbf r}|t_i\bigg] + \mathcal{O}(\Lambda t^2),
\end{equation}
where we adopt our former convention of $r_s = $ constant. Furthermore, for the uplink,
\begin{equation}
r_s =\left\{\begin{array}{l l} r_1 & \quad r_2 > r_1 \\ r_2 & \quad r_2 < r_1 \\ \end{array}\right.,
\end{equation}
and the analogous situation with $r_s$ results for the downlink, although one need not worry too much about this since these constant terms cancel in our two-way DSN frequency shift expression. Plugging in above and Taylor expanding our fractions to first-order in $(v/c)$ and $\sqrt{\Lambda}t$,
\begin{eqnarray}
\bigg[\frac{\lambda_1}{\lambda_3} \bigg]_{\rm DSN} = \bigg[1 + \sqrt{\Lambda}|\hat{\mathbf k}_1 \mathbf{\cdot} \hat{\mathbf r}| (t_3 - t_1) \bigg] \bigg[1 + 2(\sum_i v^i c_i/c^2)|_1 - 2(\sum_i v^i c_i/c^2)|_2 \bigg] \nonumber
\\ + \mathcal{O}[\Lambda t^2, (v/c)^2] \nonumber \\
= \bigg[1 + \sqrt{\Lambda}|\hat{\mathbf k}_1 \mathbf{\cdot} \hat{\mathbf r}| (t_3 - t_1) \bigg] \bigg[1 + 2({{\mathbf v}}_1 \mathbf{\cdot} \hat{\mathbf k}_1/c_1) - 2({{\mathbf v}}_2 \mathbf{\cdot} \hat{\mathbf k}_1/c_2) \bigg] + \mathcal{O}[\Lambda t^2, (v/c)^2],
\end{eqnarray}
where we have taken the spatial hypersurface of the metric in our local region of the Milky Way within our solar system to be approximately Minkowskian resulting in $\sum_i {v}^i c_i$ being equal to the ordinary Euclidean dot product ${\mathbf v} \mathbf{\cdot} c\hat{\mathbf k}$. In other words, the difference in expectation for what constitutes a physical clock-rate does not affect the Doppler terms to first-order in $\sqrt{\Lambda}t$. To see why we can approximate in this manner, we notice that the difference between the spatial portion of the Minkowski metric transformed into spherical coordinates and that of (\ref{metric}) lies in the $\cosh(\sqrt{\Lambda}t)$ term. Yet, since we are examining up to only first-order in $\sqrt{\Lambda}t$ after replacing $t \rightarrow T$ for DSN frequency calculations, these two different spatial hypersurfaces will appear to be the same. Furthermore, since all of our calculations are for spacecrafts traveling near the Sun relative to the distance to our galactic center, we take $(v/c_1) \approx (v/c_2)$ as we'll have $r_1/r_2$ multiplying factors of $v/c$ in this expression to relate terms with $c_1 = \sqrt{\Lambda}r_1$ to those with $c_2 = \sqrt{\Lambda}r_2$. Since $(v/c)$ is already small for our slow-moving massive objects, this approximation appears valid. Our general expression for the DSN-calculated two-way shift in wavelength according to the theory of inertial centers is given by
\begin{eqnarray}
\bigg[\frac{\lambda_1}{\lambda_3}\bigg]_{\rm DSN} = 1 - \frac{2\hat{\mathbf k}_1 \mathbf{\cdot} ({\mathbf v}_2 - {\mathbf v}_1)}{c_1} + 2\sqrt{\Lambda}|\hat{\mathbf k}_1 \mathbf{\cdot} \hat{\mathbf r}|(t_2 - t_1) \nonumber
\\ - \frac{4\sqrt{\Lambda}|\hat{\mathbf k}_1 \mathbf{\cdot} \hat{\mathbf r}| \hat{\mathbf k}_1 \mathbf{\cdot} ({\mathbf v}_2 - {\mathbf v}_1)}{c_1}(t_2 - t_1) + \mathcal{O}[\Lambda t^2, (v/c)^2],
\end{eqnarray}
where we have used the approximation $t_3 - t_1 = 2(t_2 - t_1)$ for electromagnetic signals traveling within our localized region of the Milky Way. For this DSN two-way wavelength shift, we find our stationary clock-drift which we derived in our discussion, $2\sqrt{\Lambda}|\hat{\mathbf k}_1 \mathbf{\cdot} \hat{\mathbf r}|(t_2 - t_1)$, in addition to the DSN-assumed expression $[\lambda_1/\lambda_3]_{\rm DSN} = 1 - 2\hat{\mathbf k}_{1} \mathbf{\cdot} ({\mathbf v}_2 - {\mathbf v}_1)/c_1$, as well as another term $-4\sqrt{\Lambda}|\hat{\mathbf k}_1 \mathbf{\cdot} \hat{\mathbf r}| \hat{\mathbf k}_1 \mathbf{\cdot} ({\mathbf v}_2 - {\mathbf v}_1)(t_2 - t_1)/c_1$ dependent upon the product of the line-of-sight velocity of the spacecraft and our stationary clock-drift term.

It is interesting to note that if we were to interpret the cause of this last term in the DSN Doppler shift as the result of an additional relative `acceleration' acting between our two observers, we would likely assume this `acceleration' to take the form
\begin{equation}\label{dopplerandtimeshift}
{\mathbf a}_{\mathrm{pert}} = 2 |\hat{\mathbf k}_1 \mathbf{\cdot} \hat{\mathbf r}|\sqrt{\Lambda} {\mathbf v}_R,
\end{equation}
where ${\mathbf v}_R$ is the line-of-sight velocity between our observers. Here, we've used the assumed form of residual DSN Doppler terms due to an unmodeled acceleration, $\mathbf{a}_{\mathrm{pert}}$,
\begin{equation}
\frac{\nu_3}{\nu_1}\bigg|_{\mathrm{observed}} - \frac{\nu_3}{\nu_1}\bigg|_{\mathrm{model}} =  -\frac{2\mathbf{a}_{\mathrm{pert}}\mathbf{\cdot} \hat{\mathbf R}}{c} (t_2 - t_1),
\end{equation}
where $\hat{\mathbf R} = \hat{\mathbf k}_1$ is the line-of-sight unit vector. Notice that (\ref{dopplerandtimeshift}) is of a similar form as the empirical solution proposed in Ref.~\refcite{IorioAULunar} to resolve the anomalous secular increases of the astronomical unit and the eccentricity of the lunar orbit
\begin{equation}
{\mathbf a}_{\mathrm{Iorio}} = kH_0 {\mathbf v}_R,
\end{equation}
where, in our case, $k =2 |\hat{\mathbf k}_1 \mathbf{\cdot} \hat{\mathbf r}|$ is similar in value to Iorio's empirical solution of $2.5 \leq k \leq 5$. Of course, in our notation, $\sqrt{\Lambda} = H_0$ where $H_0$ is the Hubble constant.

Furthermore, we wish to emphasize that this `frequency shift' is the result of the DSN coordinate parametrization of time and {\it not} a physical shift in frequency. To see this, we remark that the calculation for the actual shift in frequency according to this theory requires that we use coordinate time $t$ if we assume that coordinate time corresponds with the physical clock of a stationary observer. Returning to our general expression (\ref{genshift}) for the fractional shift in wavelength according to the theory of inertial centers, we realize that coordinate time $t$ for the initial emitter moving with the same velocity $v^i|_1 = v^i|_3$ and remaining the same distance away from an inertial center $r_1 = r_3$ has the same relation with proper time $\tau$ for emission and reception such that $(dt/d\tau)|_1 = (dt/d\tau)|_3$. Then, to first-order in $\sqrt{\Lambda}t$ and $(v/c)$, one has for the two-way fractional shift in wavelength after factoring in the relative velocities of our two observers,
\begin{equation}
\frac{\lambda_1}{\lambda_3} =1 - 2\bigg(\sum_i \frac{v^i c_i}{c^2} \bigg|_2 - \sum_i \frac{v^i c_i}{c^2}\bigg|_1 \bigg)  + \mathcal{O}[\Lambda t^2, (v/c)^2],
\end{equation}
where we obtain the DSN-assumed expression for $r_1 \approx r_2$.

\end{document}